\documentclass[a4paper,11pt]{article}
\usepackage{jheppub}
\usepackage[T1]{fontenc} % 

\usepackage[utf8]{inputenc}

\usepackage{graphicx} % Required for inserting images
\usepackage{xcolor}
\usepackage{hyperref}
\usepackage{dcolumn}% Align table columns on decimal point
\usepackage{bm}% bold math
\usepackage{amsmath,amsfonts,amssymb}
\usepackage{subcaption}
\usepackage{slashed}
\usepackage[shortlabels]{enumitem}
\usepackage[normalem]{ulem}
\usepackage{placeins}

\newcommand{\lcdm}{$\Lambda$CDM~}

\newcommand{\mpl}{M_{\mathrm{Pl}}}
\newcommand{\fpl}{f_{\mathrm{Pl}}}

\title{A Natural Dark Matter–Dark Energy Interaction from Axion–Monopole Dynamics}

\author[a,b,c,d]{Hengameh Bagherian}
\author[e,f]{Matthew Reece}
\author[a,b,c,d,g]{Lian-Tao Wang}
\author[e,f,h]{Huangyu Xiao}
\affiliation[a]{Department of Physics, University of Chicago, Chicago, IL, 60637, USA}
\affiliation[b]{Enrico Fermi Institute, The University of Chicago, Chicago, IL 60637, USA}
\affiliation[c]{Kavli Institute for Cosmological Physics, University of Chicago, Chicago, IL 60637, USA}
\affiliation[d]{Leinweber Institute for Theoretical Physics, University of Chicago, Chicago, IL 60637, USA}

\affiliation[e]{Leinweber Institute for Theoretical Physics at Harvard, Harvard University, Cambridge,
MA 02138, USA}
\affiliation[f]{Department of Physics, Harvard University, Cambridge, MA, 02138, USA}

\affiliation[g]{HEP Division, Argonne National Laboratory, 9700 Cass Ave., Argonne, IL 60439, USA
}

\affiliation[h]{Physics Department, Boston University, Boston, MA 02215, USA}

\emailAdd{hengameh@uchicago.edu}
\emailAdd{mreece@g.harvard.edu}
\emailAdd{liantaow@uchicago.edu}
\emailAdd{huangyuxiao@fas.harvard.edu}

\abstract{
We propose a natural framework for interacting dark matter and dark energy based on axion–monopole dynamics in a dark gauge sector. Dark matter is composed of dark ’t Hooft–Polyakov monopoles, while dark energy is an ultralight axion whose coupling to the dark gauge field induces an axion-dependent dyon mass through the Witten effect. This generates a direct dark matter–dark energy interaction. We show that generic models with a significant dark matter–dark energy coupling suffer from a naturalness problem: the Coleman–Weinberg potential induced by the interaction is typically much larger than the dark-energy scale. In our model, the corrections arise from a monodromy tower of dyon states, allowing them to be exponentially suppressed. In the minimal setup, the interaction strength is predicted to be $\sim 0.1\%$, below the sensitivity of current observations. We discuss scenarios in which cosmological population of higher-charge dyons can enhance the interaction to the percent level, producing an apparent phantom dark-energy equation of state. We further show that the parameter space of the axion-monopole model preferred by fitting the Dark Energy Spectroscopic Instrument (DESI) results can also naturally explain the cosmic birefringence with a small anomaly coefficient. 
}

\begin{document}
\maketitle

\section{Introduction}

According to the \lcdm fit to Planck data, only about $5\%$ of the energy content of the Universe is directly accessible through non-gravitational observations and experiments. The remaining $95\%$ consists of dark matter and dark energy, whose microscopic origin remains one of the main open problems in fundamental physics \cite{Planck:2019nip}.
While the \lcdm framework provides an excellent fit to a wide range of cosmological observations, a growing number of tensions and anomalies suggest possible deviations from this minimal picture. Most recently, results from the Dark Energy Spectroscopic Instrument (DESI) have provided intriguing hints that the dark energy sector may be dynamical or interacting rather than a strict cosmological constant \cite{DESI:2025zgx}. 
This motivates a broad class of interacting dark sector models in which dark matter and dark energy are dynamically coupled \cite{Li2026,Shah2025,Silva2025,Chakraborty2025,You2025,Petri2026}. A common realization involves a dark matter species $\chi$ whose mass, $m_{\chi}(\varphi)$, depends on a light scalar field $\varphi$ evolving in a bare potential $V(\varphi)$ \cite{Amendola:1999er,Pourtsidou:2013nha}. In such constructions, the dark matter energy density contributes an effective term to the scalar potential. At high redshift, this contribution typically dominates over $V(\varphi)$, while at late times, as the dark matter dilutes, the dynamics transition to a regime governed primarily by $V(\varphi)$. This time dependence can give rise to an effective dark energy sector capable of mimicking phantom-like behavior \cite{Das:2005yj,Aboubrahim:2024spa,Caldwell:2025inn,Liu:2025bss,Barman:2025ryg,Chen:2025ywv,Nakagawa:2025ejs,SevillanoMunoz:2026jgk,Antusch:2026ldp,Sabogal:2026bsw}.
However, this mechanism generically suffers from a severe radiative stability problem. The large hierarchy between the mass of dark matter and dark energy particles induces Coleman--Weinberg loop corrections that tend to drive the scalar mass well above its required tiny value, thereby destabilizing the flatness of the potential \cite{DAmico:2016jbm}. As a result, maintaining a technically natural and stable realization of such interacting dark sector scenarios remains a central model-building challenge.

In this work, we focus on this challenge, namely, the difficulty of constructing radiatively stable interacting dark sector models capable of producing DESI-motivated dynamical dark energy behavior. We emphasize that only a small class of mechanisms are known to address this issue in a controlled way. 
We propose a minimal realization of the latter idea based on an axion monodromy framework. In this setup, ultralight axions with masses $m_\varphi < H_0 \simeq 10^{-33}\,\mathrm{eV}$ are used to model dynamical dark energy. Axions generically couple to gauge-field topological terms and therefore provide a natural interaction with topological defects such as magnetic monopoles \cite{Fischler:1983sc,Kawasaki:2015lpf,Banerjee:2024ykz,Jeong:2023gjc,Komiya:2026uhc}. In the presence of an axion background, a monopole acquires an induced electric charge via the Witten effect and becomes a dyon \cite{Witten:1979ey}. This generates a direct coupling between axion dark energy and monopole dark matter.

We consider the case in which these monopoles reside in a dark sector and constitute the dominant component of dark matter. This is phenomenologically well motivated: ordinary GUT-scale monopoles lead to the overclosure problem in the Universe and are subject to stringent observational constraints from their magnetic charge~\cite{Zeldovich:1978wj,Patrizii:2015uea}. Therefore, GUT monopoles with $U(1)_{\rm EM}$ charge cannot account for the dark matter abundance in standard scenarios, whereas dark monopoles evade these constraints while preserving the same topological coupling structure.

This axion--monopole interaction is also favorable from the perspective of radiative stability. In such systems, the effective potential is not generated by a single matter loop but receives contributions from an infinite tower of dyonic states~\cite{Fan:2021ntg}. In the axion-monodromy effective description, this tower can soften otherwise dangerous Coleman--Weinberg corrections, placing these models among a restricted class of interacting dark sector theories that may remain radiatively controlled.
Nevertheless, in the minimal realization the axion--monopole coupling is insufficient to generate the interaction strength required to reproduce the DESI-motivated phantom behavior. We therefore also consider extensions of the minimal setup that enhance the effective coupling through populating higher $n$ charge dyons in a charge asymmetric way. 
This highlights a more general model-building tension, namely that even when a radiatively stable interacting dark sector can be constructed, there is no guarantee that it produces an effective interaction strength significant enough to account for current observational hints.

Additionally, if the dark energy axion couples to visible photons, it leads to additional observable signatures known as birefringence. Analyses of CMB polarization data suggest a nonzero isotropic cosmic birefringence angle~\cite{AtacamaCosmologyTelescope:2025blo, Eskilt:2022cff, Eskilt:2022wav, Diego-Palazuelos:2022dsq, Minami:2020odp}. Such an effect can arise from the coupling of the axion to the electromagnetic topological term, which induces a rotation of CMB polarization sourced by an axion background. Current analyses report a mild preference for a nonzero signal at the level of approximately $3\sigma$, although forthcoming CMB experiments will be required to clarify its statistical significance. Explaining this signal with axion models, however, requires enhanced couplings to photons compared to the axion decay constant or nontrivial axion dynamics \cite{Fujita:2020ecn,Nakagawa:2021nme,Co:2026rga}. We will show that the axion-monopole interactions can also naturally explain the birefringence signal without introducing large anomaly coefficients on the topological term. Moreover, the parameter space preferred by birefringence also provides a better fit to DESI data, since both observations prefer a strong coupling between dark matter and the dark energy field.

The remainder of this paper is organized as follows: In Sec.~\ref{sec:naturalDMDE}, we discuss the naturalness problem of DM--DE interaction and present a general class of solutions from a monodromy tower. In Sec.~\ref{sec: modelsetup}, we discuss our model setup that realizes the dark magnetic monopole as dark matter and the axion as dark energy with an exponentially suppressed potential. In Sec.~\ref{sec:cosmic_evol}, we study the cosmological evolution of monopoles and compute their relic density. In Sec.~\ref{sec:pheno_signature}, we discuss phenomenological signatures of our model.

\section{Natural dark matter--dark energy interactions from monodromy} \label{sec:naturalDMDE}

\subsection{A naturalness problem for dark matter--dark energy interactions}
\label{subsec:naturalnessproblem}

A simple scenario to consider is that dark energy arises from the potential energy of a scalar field $\varphi$, and dark matter is a particle $\chi$ with a mass $m_\chi(\varphi)$ that depends on the value of the scalar field. The energy density of dark matter is $\rho_\chi(\varphi) = m_\chi(\varphi) n_\chi$ with $n_\chi$ the number density of $\chi$ particles. Because $\rho_\chi(\varphi)$ is $\varphi$-dependent, this can also be viewed as a density-dependent correction to the effective potential of $\varphi$. The evolution of dark matter, dark energy, and density perturbations in such a scenario has recently been studied in~\cite{LaPenna:2026avs}.

In quantum field theory, such a scenario generically suffers from a fine-tuning problem. More specifically, we will argue the following:

\paragraph{Fine-tuning of dark matter--dark energy EFT.} Given an effective field theory of a single massive dark matter particle $\chi$ interacting with a light dark energy scalar field $\varphi$ with a cosmological epoch in which
\begin{itemize}
    \item the EFT is weakly coupled,
    \item $\chi$ behaves as non-relativistic dark matter, and
    \item the part of the $\varphi$ potential that depends on the density of $\chi$ dominates the dynamics
\end{itemize}
then (at least) one of the following must be the case:
\begin{enumerate}[(a)]
\item The $\varphi$ potential is fine-tuned.
\item $\chi$ is a light boson, in the ``wave dark matter'' regime (mass below an eV).
\item The Coleman-Weinberg estimate of the $\varphi$ potential from $\chi$ alone is a bad approximation. The theory has some underlying structure that enforces cancellations without fine-tuning.
\end{enumerate}
Furthermore, we have found that case (b) is difficult to realize in a concrete model, as discussed in Appendix~\ref{app:waveDMDE}. 

A similar argument, focused on the Coleman-Weinberg correction from loops of $\chi$ particles to the mass of $\varphi$, was already made in~\cite{DAmico:2016jbm}. We focus on the correction to the gradient of the potential and quantify when the $\chi$-dependent part of the $\varphi$ potential dominates, but the underlying idea is the same. Case (c), the focus of our work, was not considered in~\cite{DAmico:2016jbm}.

\medskip

The basic structure of the interaction between $\varphi$ and $\chi$ can be realized both for bosonic and for fermionic $\chi$, with effective Lagrangians:
\begin{align} \label{eq:Lb_main}
    {\cal L}_{\mathrm{b}} &= -\frac{1}{2} (\partial \varphi)^2 - \frac{1}{2} (\partial \chi)^2 - V(\varphi) - \frac{1}{2} m_\chi(\varphi)^2 \chi^2,\\  \label{eq:Lf_main}
    {\cal L}_{\mathrm{f}} &= -\frac{1}{2}(\partial \varphi)^2 + i \overline{\chi} \slashed{\partial} \chi - V(\varphi)  - m_\chi(\varphi) \overline{\chi}{\chi}.
\end{align}
In the fermionic case we have assumed that $\chi$ is a Dirac fermion, though the conclusions we draw would also hold for a Majorana fermion (up to factors of 2).

In order for $\chi$ to behave as dark matter and $\varphi$ as dark energy, the effective mass of $\chi$ must be much larger than that of $\varphi$. This raises a question of naturalness: does the effective interaction generate quantum corrections to the potential of $\varphi$ that are much larger than the potential itself? The 1-loop correction to the effective potential of $\varphi$ from loops of $\chi$ particles is given by the well-known Coleman-Weinberg result~\cite{Coleman:1973jx}. The 1-loop effective potential is quadratically divergent in the UV cutoff $\Lambda$ (which should be much larger than $m_\chi(\varphi)$). The coefficient of the $\Lambda^2$ term is scheme-dependent, but we can consider the simple scheme in which we add up the $\pm \frac{1}{2} \omega_{\vec k}$ zero-point energies of all $\chi$ momentum modes up to $|{\vec k}| = \Lambda$:
\begin{equation}
    V_\varphi^{(1)}(\varphi) = \frac{c_\chi}{2} \int_0^\Lambda \frac{\mathrm{d}^3{\vec k}}{(2\pi)^3} \sqrt{{\vec k}^2 + m_\chi(\varphi)^2},
\end{equation}
where $c_\chi$ is a signed count of degrees of freedom ($+1$ for $\chi$ a real scalar, $-4$ for $\chi$ a Dirac fermion). 

The classical evolution of $\varphi$ is determined by the derivative $V'(\varphi)$, so we take our naturalness criterion to be that the derivative of $V_\varphi^{(1)}(\varphi)$ should be much smaller (in magnitude) than the derivative of the tree level potential. For the loop-corrected potential, we have
\begin{align}
    \frac{\mathrm{d}}{\mathrm{d}\varphi} V_\varphi^{(1)}(\varphi) &= \frac{c_\chi}{2} m_\chi(\varphi) m_\chi'(\varphi) \int_0^\Lambda \frac{\mathrm{d}^3{\vec k}}{(2\pi)^3} \frac{1}{\sqrt{{\vec k}^2 + m_\chi(\varphi)^2}} \nonumber \\
    &= \frac{c_\chi}{16 \pi^2} m_\chi(\varphi) m_\chi'(\varphi) \left(2 \Lambda^2 - m_\chi(\varphi)^2 \log \frac{\Lambda^2}{m_\chi(\varphi)^2} + \cdots \right).
\end{align}
Here the $\cdots$ correspond to terms that are finite in the $\Lambda \to \infty$ limit. We take our naturalness criterion, then, to be
\begin{equation} \label{eq:naturalness}
|V'(\varphi)| \gg \left| \frac{\mathrm{d}}{\mathrm{d}\varphi} V_\varphi^{(1)}(\varphi)\right| \approx \left|\frac{c_\chi m_\chi(\varphi)m_\chi'(\varphi) \Lambda^2}{8\pi^2}\right|.
\end{equation}
One could reasonably ask at what value of $\varphi$ we should evaluate this condition; in particular, we should not expect it to hold at a critical point of the tree-level potential where $V'(\varphi) = 0$. However, $\varphi$ is changing with time, so even if $V'(\varphi) = 0$ at some instant, it will not be true at a somewhat earlier or later time, when we can apply the argument. If we estimate $H \dot \varphi \sim m'_{\chi} (\varphi) n_\chi$, the change in $\varphi$ in one Hubble time during the matter-dominated era can be estimated as 
\begin{equation}
\Delta \varphi \sim \frac{m'_{\chi} (\varphi) n_\chi}{H^2}  \sim \frac{\mpl^2 m'_{\chi} (\varphi) n_\chi}{H^2 \mpl^2} \sim  \frac{\mpl^2m'_{\chi} (\varphi) n_\chi}{n_\chi m_\chi(\varphi)} \sim \mpl^2 \frac{m'_{\chi} (\varphi)}{m_\chi(\varphi)}, \label{eq:Deltaphiinhubble}
\end{equation}
large enough that the possible presence of critical points is not a significant loophole in our argument. In particular, one can use this estimate to replace $V'(\varphi)$ in Eq.~\eqref{eq:naturalness} with $V''(\varphi)\Delta \varphi$, and check that the conclusion survives.

Next we turn to the criterion that, in some cosmological period of interest, the part of the $\varphi$ potential that depends on the density of $\chi$ should dominate the dynamics. The effective potential for $\varphi$ induced by the finite density of $\chi$ is given by
\begin{equation} \label{eq:Vinduced}
   V_\mathrm{ind}(\varphi) = \rho_\chi(\varphi) =  m_\chi(\varphi) n_\chi,
\end{equation}
where $n_\chi$ is the number density of non-relativistic $\chi$ particles in the universe. Requiring that this dominates the dynamics means that its derivative is larger (in magnitude) than that of the tree-level, $\chi$-independent potential:
\begin{equation} \label{eq:largedensitydependence}
    |V_\mathrm{ind}'(\varphi)| = \left| m_\chi'(\varphi) n_\chi \right| \gg |V'(\varphi)|.
\end{equation}

Chaining together Eqs.~\eqref{eq:largedensitydependence} and~\eqref{eq:naturalness}, we learn that 
\begin{equation} \label{eq:nchilarge}
    n_\chi \gg \frac{|c_\chi|}{8\pi^2} m_\chi \Lambda^2.
\end{equation}
Because $\Lambda \gg m_\chi(\varphi)$, this implies a large phase-space density of particles. In particular, it is incompatible with fermionic $\chi$, because it leads to a  Fermi energy
\begin{equation}
    E_F = \frac{1}{2m_\chi} (3 \pi^2 n_\chi)^{2/3}  \gg 0.66 \left(\frac{\Lambda}{m_\chi}\right)^{4/3}m_\chi \gtrsim m_\chi.
\end{equation}
In order to treat the dark matter as non-relativistic, we require $E_F \lesssim m_\chi$. In other words: there is no regime in which the $\varphi$ potential is natural and determined by the Coleman-Weinberg estimate, the $\chi$ particles are non-relativistic fermionic dark matter, {\em and} the density-dependent $\varphi$ potential dominates over the tree-level $\varphi$ potential.

In the bosonic case, there is not a sharp contradiction, but Eq.~\eqref{eq:nchilarge} implies a stringent upper bound on the mass of $\chi$. In particular, at the time of matter-radiation equality, the CMB temperature was $T_\mathrm{eq} \approx 0.8\,\mathrm{eV}$ and so we have
\begin{equation}
    m_\chi n_\chi = \frac{\pi^2}{30} g_* T_\mathrm{eq}^4 \approx (0.8\,\mathrm{eV})^4.
\end{equation}
Together with Eq.~\eqref{eq:nchilarge}, this implies that \begin{equation}
m_\chi \ll 1.7\,\mathrm{eV}.
\end{equation} 
This is characteristic of light bosonic dark matter in the ``wave dark matter'' regime, where inside galaxies it is better modeled as a classical field than as a collection of particles (see~\cite{Hui:2021tkt} for a review with further references). Fermionic dark matter in this mass range is excluded by the Tremaine-Gunn bound~\cite{Tremaine:1979we} (essentially the same Pauli exclusion argument indicated above).

This argument establishes our claim: either our use of the Coleman-Weinberg formula is misleading (either because of fine-tuning, or a physical mechanism for a cancellation) {\em or} $\chi$ is a light boson. It would be interesting to further explore the possibility that dark matter and dark energy are both light bosons, e.g., axion-like particles. In this case it is still a challenge to provide a UV origin of the very large separation of scales between the $\varphi$ and $\chi$ masses, while having a strong enough interaction between them to be cosmologically interesting. As we explain in Appendix~\ref{app:waveDMDE}, the simplest attempt to build such a model runs into a {\em tree-level} fine-tuning problem. (A theory with {\em three} fields does better~\cite{DAmico:2016jbm}.) The next question is: what mechanism could cancel the Coleman-Weinberg potential? This is also nontrivial. Supersymmetry, for example, effectively replaces the cutoff $\Lambda$ with the mass scale of the superpartner of $\chi$, if this superpartner is heavier than $\chi$ itself. However, the above argument works even when $\Lambda \approx m_\chi$, so supersymmetry is of no use here. Instead, we will consider the case where $\chi$ is part of an infinite family of states coupled to $\varphi$, known as a {\em monodromy tower}. In this case, it is known that summing up the loop corrections of the entire tower can give a result that is exponentially smaller than the contribution of a single state.

Before elaborating on our specific mechanism, let us mention two other recent theoretically motivated models of DM--DE interactions that rely on monodromy or towers of new fields. These each have some features in common with our discussion, but differ in other respects.

The fading dark sector scenario studied in~\cite{Agrawal:2019dlm,Bedroya:2025fwh} is motivated by the observation that in quantum gravity the cutoff scale decays exponentially with field space distance~\cite{Ooguri:2006in}. Assuming that dark energy is driven by a modulus field rolling toward asymptotically large values, this leads to a characteristic correlation between time-dependent dark energy and a time-dependent dark matter mass. In this case, it is the dark matter mass itself, rather than a correction to it, that decays over time as the dark energy field evolves, which can lead to large expected effects. This scenario appeals to conjectured scaling laws in quantum gravity to explain the hierarchy between the dark matter mass and the Hubble scale.

In~\cite{Khoury:2025txd,Delaunay:2026jto,Delaunay:2026fse,Khoury:2026svx}, an axion potential generated by coupling to a dark QCD sector provides the DE, while dark matter is assumed to be dark baryons. The natural size of the axion potential is set by the QCD scale, which for a dark energy axion implies a dark baryon mass much too light to constitute non-relativistic dark matter at the time the CMB formed. An extra suppression factor $\epsilon$ to make the zero-density axion potential much smaller than the expectation from dark QCD is thus needed. On the other hand, DM--DE interactions require the finite-density axion potential generated by dark baryons to be unsuppressed. There is precisely one known model that achieves this effect: the $\mathbb{Z}_N$-axion model of~\cite{Hook:2018jle}. This model contains $N$ copies of SU(3)$_C$ gluons, all coupled to a single axion, with staggered $\theta$ terms: the $k$th copy has a theta term $\varphi(x) + \frac{2\pi k}{N}$. This structure can be enforced by a monodromy under which a fractional axion shift $\varphi(x) \mapsto \varphi(x) + 2\pi/N$ is accompanied by a cyclic permutation of the $N$ gluon fields. Destructive interference between contributions from the $N$ sectors leads to an axion potential suppressed by the exponentially small factor $\epsilon \sim z^{-N}$ where $z \sim m_d/m_u > 1$ in a theory with two light quark flavors $u, d$. On the other hand, if only a single one of the $N$ sectors is populated, the finite-density potential from dark baryons will not cancel and gives an unsuppressed DM--DE interaction. This is an interesting model where a monodromy acts on a finite number of states, leading to an exponentially suppressed potential. 

We will now turn to our scenario, in which a monodromy acts on an infinite number of states, leading again to an exponentially small potential.

\subsection{A naturally small potential from a monodromy tower}

Here, we briefly review the way in which a monodromy tower of $\chi$ states can generate an exponentially small potential for $\varphi$. We assume that $\varphi$ is a periodic field, $\varphi \cong \varphi + 2\pi f$ for some decay constant $f$. The periodicity of $\varphi$ need not be trivial: it can map the theory back to itself in a nontrivial way. For example, in SU($n$) gauge theory, physics is $2\pi$ periodic in the $\theta$ term in the nontrivial sense that the Hamiltonians at $\theta$ and $\theta + 2\pi$ are not {\em equal} but are unitarily equivalent, $H_{\theta+2\pi} = UH_\theta U^\dagger$, where $U = \exp\left(\frac{i}{4\pi}\int d^3x\,\mathrm{tr}(K_\mathrm{CS})\right)$ is a Chern-Simons operator.

We assume that the periodicity of $\varphi$ is realized nontrivially through a monodromy tower. The monodromy tower is an infinite family of fields $\chi_n$ (labeled by integers $n \in \mathbb{Z}$), each of which has a mass $m_n(\varphi)$ that depends on the value of $\varphi$:
\begin{equation}
    m_n(\varphi)^2 \approx m_\chi^2 + m_{\chi \varphi}^2 \left(n - \frac{\varphi}{2\pi f}\right)^2 + \cdots, 
\end{equation}
where the $\cdots$ allow for higher powers of $(n - \varphi/2\pi f)$. Under the gauge transformation $\varphi \mapsto \varphi + 2\pi f$, the mass of an individual state is not invariant, but the entire spectrum is: the mass of the state that was originally labeled by $n$ is now the mass of the state labeled by $n+1$. This rearrangement of the states is referred to as monodromy. The effective potential for $\varphi$ generated by integrating out the entire tower of $\chi_n$ states can naturally be exponentially small. This is a calculation that has appeared in the literature in different contexts, from integrating out Kaluza-Klein modes in an extra dimension~\cite{Hosotani:1983xw, Cheng:2002iz, Arkani-Hamed:2007ryu} to integrating out loops of magnetic monopoles~\cite{Fan:2021ntg,GarciaGarcia:2025uub}. 

Integrating out a single state $\chi_n$ generates a Coleman-Weinberg potential along the lines that we estimated above, but this potential is not compatible with the periodicity of $\varphi$. It is only when we sum over all $n$ that we find a periodic result. Mathematically, this can be accomplished by Poisson resummation, which rearranges the series as a sum over a different integer $w$, where each term is proportional to $\exp(i w \varphi/f)$ and thus manifestly periodic. Another way to think about this computation is through the first-quantized worldline formalism, where all of the $\chi_n$ particles are now unified through a compact boson $\sigma$ living on the $\chi$ worldline, with a topological coupling $\frac{1}{2\pi f} \varphi\, \mathrm{d}\sigma$ to the scalar $\varphi$~\cite{Heidenreich:2020pkc} and a kinetic term $\frac{1}{2}\frac{m_\chi}{m_{\chi \varphi}^2}{\dot \sigma}^2$. The state $\chi_n$ corresponds to a state with $n$ units of momentum for $\sigma$ in the worldline quantum mechanics. The terms $\exp(i w \varphi/f)$ then correspond to having $w$ units of {\em winding} for $\sigma$ around a closed worldline. When $m_\chi^2 \gg m_{\chi \varphi}^2$, the worldline computation has an accurate semiclassical approximation~\cite{Fan:2021ntg}, suppressed by $\exp(-S)$ where $S$ is the worldline instanton action
\begin{equation}
    S = \frac{2\pi m_\chi}{m_{\chi \varphi}}.
\end{equation}
By exponentially suppressing the size of the $\varphi$ potential induced by loops of $\chi$, the monodromy structure leads to a dramatic failure of the Coleman-Weinberg estimate and evades our earlier conclusion. In particular, the mass of $\chi$ can be far above the eV scale without any fine-tuning.

To exploit this mechanism, then, we should focus on models where $\varphi$ is a periodic (axion-like) field coupled to a monodromy tower. Quite generally, these monodromy towers arise when axions and gauge fields participate in Chern-Simons terms. When there is a term of the form
\begin{equation}
   \int k \frac{\varphi}{8\pi^2 f} \mathrm{d}A_1 \wedge \mathrm{d}A_2, \label{eq:genchernsimons}
\end{equation}
then a particle magnetically charged under $A_1$ acquires a set of excitations that are electrically charged under $A_2$ (and vice versa). This is a special case of the phenomenon of anomaly inflow~\cite{Callan:1984sa}, as emphasized in this context by~\cite{Fukuda:2020imw, Heidenreich:2020pkc, Heidenreich:2021yda}. Recently, it has also been understood that this structure relates to generalized symmetries. The Chern-Simons term breaks the shift symmetry of $\varphi$, so the usual logic of technical naturalness would lead us to expect that an axion mass is automatically generated in its presence. However, a more subtle non-invertible axion shift symmetry survives in the presence of the Chern-Simons term, and is only broken when magnetic monopoles exist~\cite{Cordova:2022ieu}. This is why loops of magnetic monopoles (and their dyonic excitations) are able to generate an axion mass, as calculated in~\cite{Fan:2021ntg}, but the Chern-Simons term alone does not lead to an axion mass.

Another example of the monodromy structure arises in Kaluza-Klein theories. If we compactify a 5d gauge theory with gauge field $A$ on a circle, then in the lower-dimensional theory we have an axion field that corresponds to the holonomy around the circle, $\varphi = \oint_{S^1} A$, as well as two gauge fields, $A$ inherited from the 5d gauge field and $B$ from the 5d graviton. The holonomy effectively modifies the periodicity of charged fields around the circle and shifts the quantization of Kaluza-Klein charge. More precisely, we impose $\chi(y) \cong \chi(y + 2\pi L)$ and so the solutions take the form $\chi(y) = \exp(\mathrm{i} n y/L)$ with $n \in \mathbb{Z}$. For charged fields $\chi$, the physical Kaluza-Klein charge comes from the covariant derivative $D_y = \partial_y + i q A_y$, which acquires a $\varphi$-dependent fractional part proportional to the ordinary electric charge, $q_\mathrm{KK} = n + q \frac{\varphi}{2\pi}$. This is very similar to the Witten effect, and associated to a similar monodromy: when $\varphi \mapsto \varphi + 2\pi$, $A \mapsto A + B$. There is a corresponding Stueckelberg structure and a modified Bianchi identity: the gauge-invariant 4d field strength is $\tilde{F} = \mathrm{d}A - \frac{\varphi}{2\pi} \mathrm{d}B$, with $\mathrm{d}\tilde{F} = -\frac{1}{2\pi} \mathrm{d}B \wedge \mathrm{d}\varphi$ (for a textbook discussion of such modified Bianchi identities, see, e.g., Sec.~12.1 of~\cite{Polchinski:1998rr}; for this particular case, see, e.g.,~\cite{Heidenreich:2015nta, Heidenreich:2021yda}). This example might, at first glance, seem different from that of Eq.~\eqref{eq:genchernsimons}, since here we have a tower of purely electrically charged states. However, the Stueckelberg structure is {\em dual} to a Chern-Simons term; the structure is identical, in an electric-magnetic dual frame.

The phenomenology that we discuss in this paper depends only on having a monodromy tower, some states of which are populated in the early universe. Because the masses of these states depend on the value of the axion $\varphi$, the cosmological population of states in the tower generates a density-dependent effective potential for $\varphi$ as desired. We will focus on the case where these states are magnetic monopoles with a Witten effect coupling to the axion, because there is a straightforward mechanism for populating such monopoles during a 4d cosmological phase transition. The case of a Kaluza-Klein monodromy tower, by contrast, would require that we populate states that are best described with a 5d effective theory, which requires a more complicated cosmology (e.g., raising questions about whether 4d inflation occurred, and if not, what is responsible for the nearly scale-invariant primordial density perturbations). It would be interesting to explore the case of a KK monodromy tower, or other examples, in future work.  

\section{Model Setup}\label{sec: modelsetup}
In our model, the magnetic monopoles with a dark charge will be dark matter, and an extremely light axion field that couples to the dark U(1) serves as dark energy. The axion field induces a dark electric charge on the monopoles via the Witten effect, changing their mass and thus introducing the DM--DE interaction. We will present the predicted strength of DM--DE interaction in our minimal model, which is about one order of magnitude weaker than what is required to explain the recent DESI experiment. In the next section, we will discuss a scenario that enhances the DM--DE interaction strength by populating higher charge dyons cosmologically.

\subsection{'t Hooft Polyakov monopoles}

Magnetic monopoles represent a theoretically compelling dark matter candidate, arising naturally in gauge theories with spontaneously broken symmetries and non-trivial vacuum topology. We assume a minimal realization of this scenario via the Higgsing of a dark non-Abelian $SU(2)'$ gauge field, $W^a_{\mu\nu}$, described by the Georgi-Glashow Lagrangian:
\begin{equation}\label{eq:GG_Lag}
    \mathcal{L} \supset -\frac{1}{4} W^a_{\mu\nu} W^{a\mu\nu} - \frac{1}{2} D_\mu \phi^a D^\mu \phi^a - V[\phi].
\end{equation}
The potential is given by $V[\phi] = \frac{\lambda}{4}(\phi^a \phi^a - v_D^2)^2$, where the dark scalar field $\phi^a$ is in the adjoint representation. This field acquires a vacuum expectation value (VEV), $\langle \phi^a \phi^a \rangle = v_D^2$, which spontaneously breaks the symmetry as $SU(2)' \to U(1)'$. In this phase, magnetic monopoles—specifically 't Hooft-Polyakov monopoles—appear as topological solitons corresponding to the homotopy classes of mappings from the sphere at spatial infinity to the vacuum manifold. These configurations are classified by the second homotopy group $\pi_2(SU(2)'/U(1)') \cong \mathbb{Z}$. In the BPS limit, the mass and magnetic charge of the monopole are determined by the symmetry breaking scale, with $m_{\rm M} = 4\pi v_D/e_D$ and $q_{\rm M} = 4\pi w/e_D$, where $w \in \mathbb{Z}$ represents the topological degree (similar to winding number) of the mapping. For the rest of this paper we take $w=1$.

By extending the Georgi-Glashow Lagrangian in Eq.~\eqref{eq:GG_Lag} to include the topological term $(\theta \alpha_D / 8\pi) W^a \tilde{W}^a$ and promoting the parameter $\theta$ to a dynamical axion field via $\theta \equiv \varphi/f_\varphi$, contributions from small instantons in the broken $SU(2)'$ sector generate a periodic potential for the axion. In this notation, we take $\alpha_D = e_D^2/4\pi$ to be the fine structure constant of the dark $U(1)'$ sector. In the broken phase, the effective Lagrangian thus takes the form:
\begin{equation}\label{eq:u1lag}
    \mathcal{L} \supset \frac{\alpha_D}{8\pi} \frac{\varphi}{f_\varphi} F^\prime \tilde{F}^\prime - \Lambda^4 \left(1 - \cos\left[\frac{\varphi}{f_\varphi}\right]\right).
\end{equation}
In the unitary gauge, where $\hat{\phi}^a = (0, 0, 1)$, the $U(1)'$ field strength is identified as $F'_{\mu\nu} = \partial_\mu W^3_\nu - \partial_\nu W^3_\mu$ far from the monopole core. The magnitude of the potential barrier, $\Lambda^4 = m_\varphi^2 f_\varphi^2$, is defined by the axion mass and the decay constant $f_\varphi$ or the scale of the global $U(1)$ symmetry breaking. This scale is exponentially suppressed by the $SU(2)'$ instanton action and is approximately given by
\begin{equation}\label{eq:lambdaFM}
    \Lambda^4 = m_\varphi^2f_\varphi^2 \sim v_D^4 e^{-2\pi/\alpha_D}.
\end{equation}
This exponential suppression naturally allows the axion potential to match the observed dark energy density, provided the symmetry breaking scale and coupling are appropriately tuned. For instance, fixing $v_D = 10$~TeV, this uniquely determines the dark gauge coupling as $\alpha_D \approx 0.044$. As we will see, this is the relevant scale to produce the correct relic abundance of dark monopoles, though the exact value of $\alpha_D$ only logarithmically depends on the scale.

\subsection{Witten effect and dark matter-dark energy interaction}\label{sec:witten-inter}

An ultralight axion field rolling in the late Universe can lead to rich cosmological signatures and novel phenomenology. Crucially, in this framework, magnetic monopoles acquire an induced electric charge in the presence of a dynamical $\theta$-background—a phenomenon known as the Witten effect~\cite{Witten:1979ey}. The topological coupling $\mathcal{L} \supset (\theta \alpha_D / 8\pi) F^\prime \tilde{F}^\prime$ from Eq.~\eqref{eq:u1lag} transforms the monopoles into \textit{dyons} with a dark electric charge given by
\begin{equation}
    q_e = e_D \left( n - \frac{\theta}{2\pi} \right),
\end{equation}
where $n \in \mathbb{Z}$ denotes the quantized contribution to the electric charge. As we will discuss, this induced charge facilitates a direct, dynamical coupling between the monopole dark matter and the axionic dark energy.

As a consequence of the Witten effect, the axion also introduces corrections to the 't Hooft-Polyakov monopole mass, $m_{\rm M}= 4\pi v_D/e_D$. This shift can be understood as the additional electromagnetic energy carried by the dyon resulting from the induced electric charge in a dynamical $\theta$-background. Consider a family of dyons labeled by an integer $n$, with mass formula
\begin{equation}\label{eq:dyon_mass_general}
    m_D(n, \theta) = m_{\rm M} + \gamma m_{\rm M} \frac{\alpha_D^2}{2} \left(n - \frac{\theta}{2\pi}\right)^2 + \cdots.
\end{equation}
Here $\gamma$ is expected to be an order-one number; in particular, $\gamma = 1$ in the BPS limit. $\gamma$ may be much smaller than 1 if we introduce extremely light fermions. For the purpose of this work, $\gamma$ is an order one number since we only consider heavy fermions in the model.
The dots indicate terms of order $(n - \theta/(2\pi))^4$ or higher, which we assume to be small. Without loss of generality, we will take $\theta$ in the range $-\pi < \theta \leq \pi$. In order to trust perturbation theory, we require
\begin{equation}
    n \lesssim \alpha_D^{-1}.
\end{equation}

The presence of magnetic monopoles back-reacts on the axion field, effectively modifying its potential. In a Universe populated by monopoles and anti-monopoles, the axion acquires an additional mass term arising from the induced electric charge density on the solitons~\cite{Fischler:1983sc}. If the axion field is ultralight and drives the late-time expansion as dark energy, this mechanism naturally introduces a sizable coupling between dark energy and dark matter, assuming the monopoles constitute the dark matter density. Since $\alpha_D \approx 0.044$ is fixed by the requirement that the exponentially suppressed instanton potential matches the observed dark energy density, there is no free parameter in this minimal model setup. 
The total axion potential, expressed in terms of the dimensionless field $\theta \equiv \varphi/f_\varphi$, is given by the sum of the instanton-induced potential and the monopole energy density contribution:
\begin{equation}\label{eq:Veff}
    V_{\rm eff}(\theta) = V_0+V_{\rm ind} \equiv \Lambda^4(1 - \cos\theta) + \gamma m_{\rm M}n_{\rm M} \frac{\alpha_D^2}{2} \left(n-\frac{\theta}{2\pi}\right)^2,
\end{equation}
where $\Lambda^4 = m_\varphi^2f_\varphi^2$ from Eq.~\eqref{eq:lambdaFM}.
\subsection{DM--DE interaction strength}
To estimate the strength of the coupling from $V_{\rm ind}$, we will suppose that a number of different dyonic states are populated. In particular, let $f_n$ denote the fraction of the total magnetic monopole dark matter number density accounted for by the dyon mode $n$, so that $\sum_{n \in \mathbb{Z}} f_n = 1$. As in Eq.~\eqref{eq:Vinduced}, we can view the dark matter density, with its dependence on $\theta$, as producing an effective potential for the field $\varphi = f_\varphi \theta$. Subtracting off the $\varphi$-independent part of the total dyon density, we obtain an effective potential given by
\begin{align}
    V_\mathrm{ind}(\varphi) &= \sum_{n \in \mathbb{Z}} (\rho_n(\varphi) - \rho_n(0)) \nonumber \\
    &= \sum_{n \in \mathbb{Z}} \frac{\gamma \alpha_D^2}{2} f_n n_{\rm M} m_{\rm M}  \left(-2n \left(\frac{\varphi}{2\pi f_\varphi}\right) + \left(\frac{\varphi}{2\pi f_\varphi}\right)^2\right).
\end{align}
If the dyon charge is symmetric, i.e., $f_n = f_{-n}$, the induced potential can be simplified as
\begin{align}
        V_\mathrm{ind}(\varphi) = n_{\rm M} m_{\rm M}  \frac{\gamma \alpha_D^2}{2} \left(\frac{\varphi}{2\pi f_\varphi}\right)^2.
\end{align}
In such cases, $V_\mathrm{ind}(\varphi)$ is always a small contribution to the overall energy budget of the universe. The induced potential is of order $\alpha_D^2 \rho_{\rm M}/2$ at $\varphi\sim 2\pi f_\varphi$, therefore the DM--DE interaction is only a correction at $\sim\alpha_D^2/2\approx 0.1\%$ level to the DM density with charge-symmetric dyons.
Alternatively, if the dyon plasma is not charge-symmetric, i.e., $f_n \neq f_{-n}$, the induced potential can be dominated by higher $n$ charges.\footnote{The universe is always charge symmetric in the sense that there are both monopoles and anti-monopoles, with the population of $(q_{\rm M}, q_e)$ states matching those of $(-q_{\rm M}, -q_e)$ states. Here we use ``charge-symmetric'' to mean $f_n = f_{-n}$, a convenient abuse of terminology since these states are not actually related by a charge conjugation symmetry.} We use the effective dyon shift $n_e\equiv\sum_{n\in\mathbb{Z}} n f_n$ to define the induced potential
\begin{align}
        V_\mathrm{ind}(\varphi) = n_{\rm M} m_{\rm M} \frac{\gamma \alpha_D^2}{2} \left( -2n_e\left(\frac{\varphi}{2\pi f_\varphi}\right)+\left(\frac{\varphi}{2\pi f_\varphi}\right)^2\right).
\end{align}
A large $n_e$ clearly suggests a stronger DM--DE interaction. We consider its value from 0 to $\sim 1/\alpha_D$.
In Sec.~\ref{sec:cosmic_evol}, we will discuss in detail how to make dyons not charge-symmetric through a rotating axion field, which can greatly enhance the DM--DE interaction strength.

\section{The cosmological evolution of magnetic monopoles}\label{sec:cosmic_evol}
Throughout this work, we have assumed that magnetic monopoles are the dominant form of dark matter. In this section, we will discuss the model space that naturally achieves the monopole dark matter scenario. Also, we discuss scenarios that realize the production of higher charge dyons in a charge asymmetric way, which can further enhance the DM--DE interaction strength.

\subsection{Relic density of monopoles}\label{sec:mon_density}
The initial density of monopoles after the phase transition depends on the exact dynamics, which can be model-dependent. However, these monopoles can further annihilate away in some mass ranges due to their magnetic forces between each other, providing a ``freeze-out'' abundance of magnetic monopoles that does not sensitively depend on the dynamics of the dark sector phase transition. Schematically, the initial density is determined by \cite{1976JPhA....9.1387K}
\begin{equation}
    n_{\rm M}(t_i)\sim A\,\xi^{-3},
\end{equation}
where $\xi$ is the Higgs field correlation length at the time of phase transition, and $A\sim 0.1$ is a geometric factor. Causality requires that the correlation length must be smaller than the horizon scale, which imposes a lower bound \cite{Einhorn:1980ik}
\begin{equation}
    n_{\rm M}(t_i)\gtrsim A\,t_i^{-3}.
\end{equation}
For the usual GUT phase transition that produces magnetic monopoles via the Kibble mechanism, the initial relic abundance will overclose the Universe and the annihilation of magnetic monopoles can not suppress the density significantly \cite{Vilenkin:2000jqa}.
However, if the phase transition occurs within a dark sector at a temperature much lower than the GUT scale, the annihilation process after the phase transition governs the final relic density and it can reproduce the correct dark matter abundance~\cite{Preskill:1979zi}. 

In the plasma with dark sector particles, monopoles and antimonopoles gradually drift towards each other due to the attractive forces between them, which enhances the annihilation rate. During the diffusion process of monopoles, the drift velocity depends on the separation between monopoles and anti-monopoles, and the balance between the attractive force $F_{\rm att}\sim q_{\rm M}^2/r^2$ and the drag force $F_{\rm drag}\sim -\beta_c T^2v$
\begin{equation}
    v (r) \sim \frac{q_{\rm M}^2}{\beta_c T^2 r^2},
\end{equation}
where $\beta_c=g_{*,D}(T) $ is the degrees of freedom in the dark sector that characterizes the drag force experienced by magnetic monopoles in plasma, and $q_{\rm M}=4\pi/e_D$ is the magnetic charge.  
By requiring the annihilation timescale $\tau_d\approx d/v(d)$ equal to the Hubble time with $d=n_{\rm M}^{-1/3}$, we obtain the following monopole number density: 
\begin{equation}
    \frac{n_{\rm M}}{s}\approx 0.76\frac{\beta_c}{g_*^{1/2}(T)q_{\rm M}^2}\frac{T}{M_{\rm Pl}}.
\end{equation}
However, this process of annihilating monopoles through the attractive forces will not reduce the monopole abundance indefinitely. The most stringent condition to satisfy is that the dark sector particles must be relativistic such that $\beta_c$ remains nonzero. Below this temperature, the annihilation ceases to be efficient due to the absence of the drag force that can dissipate the monopole kinetic energy. Therefore, the temperature when the diffusion stops is $T_c\sim \alpha_D m_{\rm M}$, and the monopole-to-entropy ratio should be evaluated at $T_c$. Different dyonic states will also be populated when they scatter in equilibrium in the plasma since their masses are comparable. Therefore, the final abundance will be multiplied by a factor of $N_M\sim 2/\alpha_D$ since we expect to populate dyonic states from $-n$ to $n$, with $n\sim 1/\alpha_D$ setting the limit where the EFT of magnetic monopoles breaks down. We finally obtain
\begin{equation}
    \frac{\rho_{\rm M}}{s}\approx 0.76 \frac{\alpha_D\beta_c}{2\pi \,{g^{1/2}_*(T_c)}}\frac{m^2_{\rm M}}{M_{\rm Pl}}\approx 0.44 \, {\rm eV} \left(\frac{\alpha_D}{0.04} \right)\left(\frac{\beta_c^2}{g_*(T_c)}\right)^{1/2} \left(\frac{m_{\rm M}}{5\times10^{2}\rm \,TeV}\right)^2.
\end{equation}
Therefore, we need a monopole mass of $\sim 10^6$ GeV to reproduce the correct dark matter abundance that satisfies $\rho_{\rm M}/s=0.44$ eV. For recent studies on the cosmological evolution of hidden-sector monopoles, see \cite{Perri:2025vtn,Liveoak:2026muy,Brummer:2026wma,Pesut:2026fyl,Zhou:2025zzz}.

A point of concern for the monopole dark matter scenario is that the monopole abundance is usually smaller than that of dark gauge bosons \cite{Baek:2013dwa,Brummer:2025inh,Xie:2026vor}. In the case that dark gauge bosons are stable, their annihilation may not reduce its abundance fast enough to avoid overclosure. This is because the annihilation between dark gauge bosons is through the dark electric coupling, which is usually much weaker than the dark magnetic coupling that governs the annihilation between dark monopoles. Therefore, the abundance of dark gauge bosons is universally higher than that of monopole in these thermal production scenarios.

In this work, we introduce dark-charged fermions in the dark sector as a minimal solution. The dark $W$ bosons will be unstable and decay to them, while these fermions will further annihilate away.\footnote{Alternatively, one could consider light scalars charged under the dark $SU(2)$ to destabilize the dark $W$ bosons. These scalars can be made extremely light without altering the Witten's effect of monopoles, and the light scalars can annihilate away with an efficient rate. This usually requires fine-tuning of the scalar mass due to the mass corrections from the $W$ bosons.}  
For fermions with masses much lighter than the dark $W$ bosons, the relic abundance can be greatly suppressed \cite{Co:2025lcs}. However, light fermions will also reduce the DM--DE interaction since it can modify the dyon mass splitting, as discussed in Appendix~\ref{app:dyon_mass_formula}. 
One approach is to consider a fermion mass that is comparable to the gauge boson mass and have dark fermions further charged under another confining dark sector, leading to a stronger annihilation. The dark radiation produced by this strong annihilation can evade bounds on $N_{\rm eff}$ if the dark sector remains sufficiently colder than the visible sector. Its contribution scales as $\Delta N_{\rm eff}\propto (T_D/T_{\rm vis})^4$, so a temperature ratio $T_D/T_{\rm vis}\sim0.1$ at the relevant epoch gives a suppression of order $10^{-4}$ for fixed relativistic particle content.

Here we assume the confinement scale is comparable to the dark fermion mass, and consider an annihilation cross-section of $\sigma_B=4\pi/m_f^2$ with $m_f\approx m_W/2= \alpha_D m_{\rm M}/2$. The freeze-out abundance of these dark fermions is thus given by
\begin{equation}
    \frac{\rho_f}{\rho_{\rm DM}}\approx 1.88\times 10^{-9}\left(\frac{\rm GeV^{-2}}{\sigma_B}\right)\approx 0.02\,\left(\frac{m_{\rm M}}{5\times10^{2}\rm \,TeV}\right)^2.
\end{equation}
Now we can compare the freeze-out abundance of these dark fermions to the final abundance of magnetic monopole dark matter.
When we plug in the monopole mass that produces the dark matter relic abundance, the dark fermion abundance is only subdominant. Therefore, heavy fermions can also be made compatible with the monopole dark matter story. As we will discuss later, this approach has the advantage that the DM--DE interaction strength can be enhanced by populating high $n$ dyons with a rotating $\theta$. 

\subsection{Enhancing the signal with higher $n_e$ dyons}
Larger values of $n_e$ strengthen the DM--DE interaction and improve agreement with DESI. Populating these higher-charge dyon states requires a period of rapid axion rotation after monopole production. Such rotation is difficult to obtain from the dark-energy field with $f_\varphi\sim\mpl$ while respecting the energy budget. For an axion rotating at its vacuum radius, the ratio of its kinetic energy to the total energy density is
\begin{equation}
    \frac{\rho_{\rm kin}}{\rho_{\rm tot}}
    =\frac{f_\varphi^2\dot{\theta}^2}{6H^2\mpl^2}.
\end{equation}
Achieving $\dot{\theta}/H\gg1$ therefore requires $f_\varphi\ll \mpl$ if the kinetic energy is to remain subdominant.

We resolve this by introducing a second axion that also couples to the dark $SU(2)$, with a decay constant and initial rotation chosen independently of the parameters governing dark energy. Denoting the dark-energy axion by $\theta=\varphi/f_\varphi$ and the second axion by $\theta'=\varphi'/f_{\varphi'}$, the angle entering the dyon mass spectrum is their sum, $\Theta=\theta+\theta'$. A displacement $\Delta\theta'$ shifts the electric-charge offset $n-\Theta/(2\pi)$ by $-\Delta\theta'/(2\pi)$, so a coherent rotation of $\theta'$ can leave the dyon population with a nonzero charge offset.

The dynamics of rotating axion fields have been studied in detail in the kinetic misalignment mechanism~\cite{Co:2019jts} and the acoustic misalignment mechanism~\cite{Bodas:2025eca,Eroncel:2025qlk}. We note that this mechanism also produces a relic density of the second axion through kinetic misalignment, which we require to be subdominant to the magnetic monopole dark matter. To describe the second axion, we write its complex PQ field as $P=(\langle S\rangle/\sqrt{2})e^{i\theta'}$, where $\langle S\rangle$ is the radial amplitude with expectation value $f_{\varphi'}$. The shift symmetry $\theta'\to\theta'+\alpha$ implies a conserved PQ charge density $n_{\theta'}=\langle S\rangle^2\dot{\theta}'$.

Initially, $\langle S\rangle$ is displaced far above $f_{\varphi'}$ in an approximately quartic potential, $V_{\rm radial}\simeq\lambda \langle S\rangle^4/4$, with effective radial mass $m_{S,\rm eff}^2 = 3\lambda \langle S\rangle^2$, sufficiently flat at large field values that Hubble friction suppresses motion. As $H$ decreases, the field begins to move at $T_{\rm osc}$, defined by $m_{S,\rm eff}(\langle S\rangle_i)\simeq 3H(T_{\rm osc})$, where $\langle S\rangle_i$ is the initial radial displacement when the radial mode starts to overcome Hubble friction. At large field values of $S$, early explicit PQ-symmetry breaking can generate an angular velocity and a charge asymmetry, as in the kinetic misalignment mechanism~\cite{Co:2019jts}. Once this source becomes negligible, the yield is approximately conserved,
\begin{equation}
    Y_{\theta'}\equiv\frac{n_{\theta'}}{s}
    =\frac{\langle S\rangle^2\dot{\theta}'}{s}
    \simeq {\rm const},
\end{equation}
where $s$ is the entropy density. This assumes conserved comoving entropy and negligible angular forces on $\theta'$ for the remainder of the cosmological evolution after charge production.

We denote the scale factor by $a$. Interactions with the thermal bath can subsequently damp radial oscillations while preserving the PQ charge, circularizing the trajectory while $\langle S\rangle \gg f_{\varphi'}$. Radial force balance gives $\langle S\rangle\dot{\theta}'^{\,2}\simeq \partial V_{\rm radial}/\partial \langle S\rangle$, which in the quartic regime implies $\dot{\theta}'\propto \langle S\rangle$. Combining with charge conservation, $\langle S\rangle^2\dot{\theta}'\propto a^{-3}$, yields $\langle S\rangle \propto a^{-1}$ and $\dot{\theta}'\propto a^{-1}$. In a radiation-dominated Universe $H\propto a^{-2}$, so the angular displacement per e-fold, $\dot{\theta}'/H = d\theta'/d\ln a$, grows as
\begin{equation}
    \frac{\dot{\theta}'}{H}\propto a.
\end{equation}
Eventually, the shrinking radial amplitude reaches $\langle S\rangle \simeq f_{\varphi'}$ and the radial mode settles at a temperature
\begin{equation}
    T_s\sim T_{\rm osc}\,\frac{f_{\varphi'}}{\langle S\rangle_i},
\end{equation}
neglecting changes in entropy degrees of freedom and details of the initial relaxation~\cite{Co:2019jts}. After settling, the radius is fixed and $\dot{\theta}'\propto a^{-3}$, so $\dot{\theta}'/H\propto a^{-1}$. Most of the angular displacement therefore accumulates near $T_s$, and monopoles must already be present during this epoch to acquire the enhanced charge offset.

We therefore choose PQ-field parameters such that $T_s\lesssim T_c\sim\alpha_Dm_{\rm M}\approx20~{\rm TeV}$, placing the main rotation around or after monopole annihilation freeze-out. To obtain an appreciable shift of the dyon mass spectrum, we take $\dot{\theta}'(T_s)/H(T_s)\sim30\pi$ as a benchmark. After this early rotation, a sufficiently massive second axion becomes trapped in its potential and oscillates about a minimum. Its equation of state then behaves as non-relativistic matter, while the original axion remains responsible for dark energy.

This relic population must remain subdominant to the monopole dark matter. At angular trapping, $|\dot{\theta}'|\sim m_{\varphi'}$, so the resulting axion number density satisfies $n_{\varphi'}\sim |n_{\theta'}^{\rm trap}|$. The pre-trapping charge yield therefore determines the late-time abundance,
\begin{equation}
    \frac{\rho_{\theta'}}{s}
    \simeq m_{\varphi'}|Y_{\theta'}|
    \sim \frac{\mathcal{N}m_{\varphi'}f_{\varphi'}^2}
              {\mpl T_c}
         \frac{\sqrt{g_*(T_c)}}{g_{*s}(T_c)},
\end{equation}
where $\mathcal{N}=\dot{\theta}'(T_s)/H(T_s)\approx 2\pi/\alpha_D$ characterizes the charge-building rotation, and we have set $T_s\simeq T_c$ in the last step. Comparing with the observed dark matter energy yield, $Y_{\rm DM}^0 \equiv \rho_{\rm DM}^0/s = 0.44~{\rm eV}$, imposes an upper bound on the decay constant,
\begin{equation}
    f_{\varphi'}\lesssim\left(\frac{g_{*s}(T_c)\,Y_{\rm DM}^0\,T_c\,\mpl}{\sqrt{g_*(T_c)}\,\mathcal{N}\,m_{\varphi'}}\right)^{1/2}\approx 4 \times 10^{13}~{\rm GeV}\left(\frac{10^{-6}~{\rm eV}}{m_{\varphi'}}\right)^{1/2}.
\end{equation}
This constraint follows from the late-Universe abundance of the second axion and is independent of the energy budget constraint discussed above, which only applies to the instantaneous axion kinetic energy.

Finally, we check whether the monopole-induced potential significantly alters the rotation of the second axion. We evaluate this effect near the radial-settling temperature $T_s$, when $\langle S\rangle\simeq f_{\varphi'}$ and the angular displacement per e-fold is largest. Although the radial mode settles at this epoch, the angular mode continues to evolve, transferring kinetic energy to the dyons as their masses increase. The rate of energy transfer from the second axion to the monopole sector is
\begin{equation}
    \left.\dot{\rho}_{\theta'}\right|_{\rm mon}\sim\alpha_D^2\rho_{\rm M} \frac{\mathcal{N}}{2\pi}\frac{\dot{\theta}'}{2\pi},
\end{equation}
where $\rho_{\rm M}$ is the monopole density, evaluated near $T_s\lesssim T_c$. Comparing the change during a Hubble time to the kinetic energy gives
\begin{equation}
    \frac{\left.\dot{\rho}_{\theta'}\right|_{\rm mon}}{H\rho_{\theta'}}\sim\frac{\alpha_D^2}{2\pi^2} \frac{\rho_{\rm M}}{\rho_R} \left(\frac{\mpl}{f_{\varphi'}}\right)^2,
\end{equation}
where we have used $3H^2\mpl^2\simeq\rho_R$ during radiation domination. Requiring this ratio to be less than unity, so that the energy-transfer timescale exceeds a Hubble time, yields a lower bound on the decay constant that is independent of the second axion mass,
\begin{equation}
    f_{\varphi'}\gtrsim \left(\frac{\alpha_D^2 Y_{\rm DM}^0}{T_s}\right)^{1/2} \mpl\approx 10^{10}\,{\rm GeV},
\end{equation}
where the numerical estimate uses $\alpha_D\simeq0.04$ and $T_s\simeq 20~{\rm TeV}$. Combined with the upper bound from the relic abundance, a viable window in $f_{\varphi'}$ exists that allows the rotation to enhance the effective dyon charge by a factor of $\alpha_D^{-1}$, increasing the DM--DE interaction strength to the few-percent level.

In the following sections, we explore cosmological signatures of this interaction in both the dark and visible sectors. We will no longer track the second axion explicitly and assume its effects are encapsulated in the effective dyon charge offset $n_e$.

\section{Phenomenological Signatures}\label{sec:pheno_signature}
The main phenomenological advantage of this model shows up when studying the recent DESI BAO results, in particular, when it comes to interpreting the apparent $w<-1$ equation of state for dark energy. We will start by laying down the background equations and explaining the dynamics of the system. Later, we will show our model provides a better fit to DESI data when the DM--DE interaction is enhanced. Lastly, we also connect our model to the cosmic birefringence effect, showing that it can accommodate a small anomaly coefficient with parameters preferred by fitting DESI data. 

\subsection{Background Equations}
We take the metric to be a flat Friedmann--Robertson--Walker background, and define the following quantities:
\begin{equation}
    x \equiv \ln a = -\ln(1+z), \quad E(x) \equiv \frac{H(x)}{H_0},\quad \fpl \equiv \frac{f_\varphi}{\mpl}, \quad
    \mu \equiv \frac{m_\varphi}{H_0}, \quad
    h \equiv \frac{H_0}{100\ {\rm km}\,{\rm s}^{-1}\,{\rm Mpc}^{-1}}.
\end{equation}
We normalize $a=1$ today. A prime in this section denotes a derivative with respect to $x$. Unless otherwise indicated, $\theta$ denotes $\theta(x)$, with $\theta_0\equiv\theta(x=0)$. When using redshift as the argument, we write $E(z)\equiv H(z)/H_0$. We denote the present critical density by $\rho_{\rm crit}\equiv3H_0^2\mpl^2$, where $\mpl=(8\pi G)^{-1/2}$ is the reduced Planck mass. Here $H_0$ is the Hubble constant of the axion--monopole cosmology, rather than a separately fixed fiducial value.  

The axion-dependent dyon mass is encoded by
\begin{equation}\label{eq:mass-modulation-background}
    A(\theta)
    \equiv 1 + \frac{\gamma\alpha_D^2}{2}
      \left(n_e-\frac{\theta}{2\pi}\right)^2
    = 1 + \frac{\gamma\alpha_D^2}{8\pi^2}
      \left(\theta-2\pi n_e\right)^2,
\end{equation}
where we set $\gamma=1$ in the numerical analysis. Note that $\theta(x)$ itself is an evolving function of redshift. The monopoles constitute the dark matter, while baryons are an uncoupled component. We additionally include a cosmological constant. 
The corresponding background densities and pressure follow
\begin{align}
    \rho_{\rm Mon}(x)
      &= \rho_{\rm crit}\,\Omega_{\rm Mon}\,e^{-3x}A(\theta), \\
    \rho_b(x)
      &= \rho_{\rm crit}\,\frac{\omega_b}{h^2}e^{-3x}, \\
    \rho_{\gamma\nu}(x)
      &= \rho_{\rm crit}\,\frac{\omega_{\gamma\nu}(x)}{h^2}, \\
    \rho_\Lambda
      &=\rho_{\rm crit}\,\Omega_\Lambda,\qquad\text{with}\quad p_\Lambda =-\rho_\Lambda.
\end{align}
As mentioned in Sec.~\ref{sec:mon_density}, as long as the dark sector remains sufficiently cold, the dark radiation contribution to the total energy is negligible. The function $\omega_{\gamma\nu}(x)$ includes photons, the effectively massless neutrino species, and one massive neutrino with
$\sum m_\nu=0.06\,\mathrm{eV}$. Explicitly, we use
\begin{align}
 \omega_{\gamma\nu}(x)
  =e^{-4x}\!\left\{\omega_\gamma
  +(N_{\rm eff}-1)\omega_{\nu}^{\rm rel}
  +\omega_{\nu}^{\rm rel}\,
   \mathcal F\!\left(\frac{m_\nu e^x}{T_{\nu0}}\right)\right\},\qquad\text{with}\quad
 \omega_{\nu}^{\rm rel}
  =\frac{7}{8}\left(\frac{4}{11}\right)^{4/3}\omega_\gamma.
 \label{eq:radiation-background}
\end{align}
The normalized Fermi--Dirac energy integral appearing here is
\begin{align}
 \mathcal F(y)
 \equiv \frac{120}{7\pi^4}\int_0^\infty
 \frac{q^2\sqrt{q^2+y^2}}{e^q+1}\,dq,
 \label{eq:fermi-dirac-integral}
\end{align}
where $q$, the dimensionless comoving neutrino momentum, is being integrated over and $T_{\nu0}$ is the present neutrino temperature. The normalization gives $\mathcal F(0)=1$. Writing the physical number density as $n_{\rm Mon}(x)=n_{\rm Mon,0}e^{-3x}$, the parameter $\Omega_{\rm Mon}\equiv n_{\rm Mon,0}m_{\rm M}/\rho_{\rm crit}$ normalizes the conserved comoving monopole abundance before multiplication by $A(\theta)$; hence the instantaneous monopole density fraction is not simply $\Omega_{\rm Mon}e^{-3x}/E^2$. Indeed, the combined quantity $\Omega_{\rm Mon}A(\theta)$ is approximately $\Omega_m$ while $A(\theta)$ is not necessarily close to 1. Finally, the axion energy density and pressure are
\begin{align}
    \rho_\theta
      &= \frac{1}{2}f_\varphi^2H^2(\theta')^2
       + m_\varphi^2f_\varphi^2\bigl[1-\cos(\theta)\bigr], \\
    p_\theta
      &= \frac{1}{2}f_\varphi^2H^2(\theta')^2
       - m_\varphi^2f_\varphi^2\bigl[1-\cos(\theta)\bigr].
\end{align}

With these definitions, the Friedmann equation and axion equation of motion used in the numerical calculation are
\begin{align}
 \left[3-\frac{1}{2}\fpl^2(\theta')^2\right]E^2(x)
 &= \mu^2\fpl^2\bigl[1-\cos(\theta)\bigr]
    +3\Omega_\Lambda \nonumber\\
 &\hspace{2em}
    +3\Omega_{\rm Mon}e^{-3x}A(\theta)
    +3\frac{\omega_b}{h^2}e^{-3x}
    +3\frac{\omega_{\gamma\nu}(x)}{h^2},
    \label{eq:background-friedmann}\\
 0= E^2\theta''
    &+\left(3E^2+\frac{1}{2}\frac{dE^2}{dx}\right)\theta' 
    +\mu^2\sin(\theta)
    +\frac{3\Omega_{\rm Mon}e^{-3x}}{\fpl^2}
      \frac{dA}{d\theta}. \label{eq:background-axion}
\end{align}

 On a final note, we retain $\Omega_\Lambda$ as a free parameter to allow the axion to constitute any fraction of the dark energy today. If instead the axion was required to account for all of it, $m_\varphi^2f_\varphi^2\sim\Omega_{\rm DE}H_0^2\mpl^2$, and the quintessence requirement $m_\varphi/H_0\sim\mathcal{O}(1)$ would fix $\fpl\sim\sqrt{\Omega_{\rm DE}}\,H_0/m_\varphi\sim\mathcal{O}(1)$. At these values the interaction term in Eq.~\eqref{eq:background-axion} and its phenomenological consequences are relatively small. Including $\Omega_\Lambda$ opens the parameter space to smaller $\fpl$, strengthening the monopole force and yielding richer phenomenology.

\subsection{The axion--monopole interaction tracks the Hubble scale}\label{sec:tracking}

\begin{figure}[t]
    \centering
    \includegraphics[width=\linewidth]
        {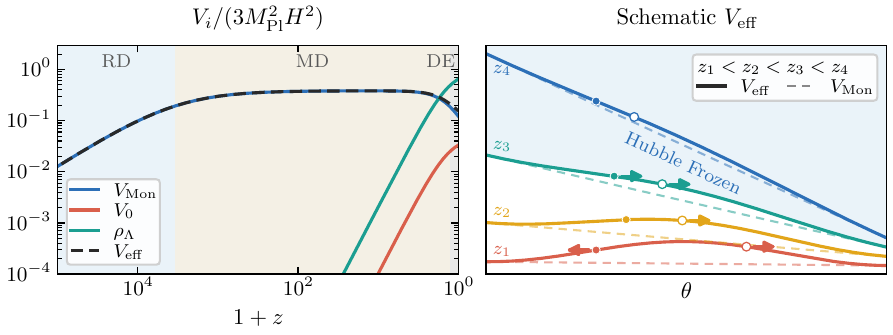}
    \caption{\textbf{Left:} $V_0$, $V_{\rm Mon}$, $V_{\rm eff}$, and $\rho_\Lambda$, each normalized by $3\mpl^2H^2$. Shading indicates radiation, matter, and dark energy domination. \textbf{Right:} Schematic evolution of $V_{\rm eff}$ (solid) and $V_{\rm Mon}$ (dashed), from high redshift (top) to the present (bottom). Filled and open markers illustrate trajectories that turn around and cross a cosine barrier, respectively. Parameters: $\Omega_\Lambda=0.65$, $\mu=0.5$, $\alpha_D=0.044$, $\theta_i=3\pi/4$, $n_e=30$, $\fpl=0.461$, $H_0=67.24\,{\rm km\,s^{-1}\,Mpc^{-1}}$, and $\Omega_{\rm Mon}=0.144$.}
    \label{fig:axion-mechanism-tracking}
\end{figure}

The axion dependence of the monopole mass gives rise to a density-dependent force on the axion field. Since this force dilutes with the monopole number density, its strength relative to Hubble friction evolves as the Universe transitions through radiation, matter, and dark energy domination. This evolution acts as a natural cosmological on/off switch for the axion--monopole coupling, with the monopole drive remaining subdominant during radiation domination, sustaining field motion throughout the matter era, and weakening again at late times as dark energy comes to dominate.

Subtracting the $\theta$-independent monopole rest mass from the effective potential yields
\begin{align}
    V_{\rm eff}(\theta,x)
    &= V_0(\theta)+V_{\rm Mon}(\theta,x),
    \label{eq:veff}\\
    V_0(\theta)
    &= m_\varphi^2f_\varphi^2[1-\cos(\theta)],\\
    V_{\rm Mon}(\theta,x)
    &= \rho_{\rm crit}\Omega_{\rm Mon}e^{-3x}
       [A(\theta)-1].
\end{align}
During radiation domination, $n_{\rm Mon}(x)/H^2\propto a$, so the monopole induced potential grows progressively stronger relative to Hubble friction. The initially frozen field can therefore begin to roll as the Universe approaches matter--radiation equality. Once matter domination is established, $n_{\rm Mon}(x)\propto a^{-3}\propto H^2$, and the ratio $V_{\rm Mon}/(3\mpl^2H^2)$ settles onto the approximate plateau visible in the left panel of Fig.~\ref{fig:axion-mechanism-tracking}. We refer to this regime as Hubble tracking.

Simultaneously, $m_\varphi\sim H_0$ ensures that the intrinsic cosine curvature remains small compared with $H^2$ throughout most of matter domination. Only at late times, as the monopole force continues to dilute while $H$ decreases more slowly, does the cosine potential regain dynamical control.

For large $n_e$, the quadratic monopole potential is centered at $\theta=2\pi n_e$ and acts approximately as a linear tilt, capable of driving the axion toward increasing $\theta$ even against the gradient of the cosine potential; see the right panel of Fig.~\ref{fig:axion-mechanism-tracking} for a schematic depiction. Crucially, this tilt can sustain appreciable field excursion during matter domination without triggering rapid oscillations. Measured against the Hubble rate, the cosine force, $\mu^2\sin\theta/E^2\propto a^3$, is suppressed during early and intermediate matter domination. This is the epoch in which the monopole-induced force dominates, as it does not redshift relative to Hubble,
\begin{equation}
    -\frac{1}{f_\varphi^2H^2}\frac{\partial V_{\rm Mon}}{\partial\theta}
    = (2\pi n_e-\theta)\frac{m_{\rm Mon}^2}{H^2},
\end{equation}
where the monopole-induced mass is
\begin{equation}
    \frac{m_{\rm Mon}^2}{H^2}
    \equiv
    \frac{1}{f_\varphi^2H^2}
    \frac{\partial^2V_{\rm Mon}}{\partial\theta^2}
    = \frac{3\alpha_D^2\Omega_{\rm Mon}e^{-3x}}
           {4\pi^2\fpl^2 E^2(x)}.
\end{equation}
Since the redshift dependence cancels during matter domination, both quantities are maintained as $H$ decreases. For the benchmark in Fig.~\ref{fig:axion-mechanism-tracking}, setting $\theta\simeq\theta_i=3\pi/4$ and neglecting the axion kinetic energy, the matter-era estimates read
\begin{equation}
    \frac{m_{\rm Mon}^2}{H^2}\simeq3.16\times10^{-4},
    \qquad
    -\frac{\partial_\theta V_{\rm Mon}}{f_\varphi^2H^2}
    \simeq5.87\times10^{-2}.
\end{equation}
Being $\mathcal{O}(\alpha_D^2)$, this force is small, and the field drifts by only $\Delta\theta\simeq\tfrac{2}{3}\times5.87\times10^{-2}\approx0.04$ per e-fold. The monopole force changes sign only at $\theta=2\pi n_e$, a distance $\simeq59\pi$ from the field range of interest, so the drift accumulated over the matter era never brings the field near this zero. The axion therefore experiences a constant, one-signed force throughout the epoch, undergoing slow, friction-damped drift rather than oscillation.

As the cosine potential regains control at late times, the displacement and velocity accumulated during the matter-era excursion determine whether the field turns around or crosses a cosine barrier, as illustrated in the right panel of Fig.~\ref{fig:axion-mechanism-tracking}. These two outcomes lead to qualitatively distinct phenomenological behaviors, discussed later in Sec.~\ref{sec:phantom_behavior}.
\subsection{Fitting Methodology}
\label{sec:desi-method}

In this section, we outline how we constrain the model parameters and select our benchmark values. The gold standard would be a full Monte Carlo analysis; however, since a complete perturbation-level treatment is beyond the scope of this paper, we instead use background-level equations to anchor the pre-recombination physics and reduce the number of free parameters. This procedure is meant to demonstrate the potential of our model for a full fit and does not replace a joint CMB+BAO likelihood analysis. A perturbative analysis with a Boltzmann solver is deferred to a follow-up paper.

For these constraint-solved backgrounds, we use the DESI-provided data vector and covariance matrix~\cite{DESI:2025zgx,DESI:2025zpo} to compute $\chi^2_{\rm BAO}$ as a measure of agreement with the data. We report the raw statistic rather than a reduced $\chi^2$. Although three parameters are determined by background constraints, the remaining inputs are scanned and multiple solutions may be selected using their BAO statistics. The effective number of fitted degrees of freedom is therefore not specified by the number of background equations alone. 

For $\alpha_D = 0.044$ and each fixed choice of $(\Omega_\Lambda,\mu,n_e,\theta_i)$, we determine $(\fpl,h,\Omega_{\rm Mon})$ from three background constraints. The axion is initialized at $z_{\rm ini}=10^{14}$ according to $\theta(x_{\rm ini})=\theta_i$ and $\theta'(x_{\rm ini})=0$; this is the initially Hubble-frozen branch of the ultralight axion. Any earlier dynamics invoked to populate higher-charge dyons are attributed to the additional axion discussed above and are not included in this late-time initial condition.

We anchor the pre-recombination sector to the CMB-only fiducial \lcdm cosmology of~\cite{Jhaveri:2026bla}, obtained from fits to ACT DR6, SPT-3G D1, and Planck 2018, with multipole cuts that limit ACT--Planck overlap, together with ACT, Planck, and SPT CMB-lensing measurements. Crucially, this reference point does not include DESI BAO. We adopt the following physical inputs and list the acoustic target calculated from our fiducial background below:
\begin{align}
 \omega_b^{\rm fid}&=0.022390, & \omega_c^{\rm fid}&=0.12034, & 100\,\theta_s^{\rm fid}&\simeq1.041441, \nonumber\\
 h^{\rm fid}&=0.6715, & r_d^{\rm fid}&=147.00\,\mathrm{Mpc}, & N_{\rm eff}&=3.044.
 \label{eq:fiducial-cosmology}
\end{align}
Here $\omega_b=\Omega_bh^2$ and $\omega_c=\Omega_ch^2$ are the physical baryon and cold-dark-matter density parameters, $\theta_s$ is the acoustic angular scale, $h=H_0/(100\,{\rm km\,s^{-1}\,Mpc^{-1}})$, $r_d$ is the comoving sound horizon at baryon drag, and $N_{\rm eff}$ is the effective number of relativistic species. The neutrino sector contains one $0.06\,\mathrm{eV}$ massive species and the remaining effectively massless contribution, as in Eq.~\eqref{eq:radiation-background}. The value of $\theta_s^{\rm fid}$ above is calculated consistently within our fixed-epoch approximation, rather than taken directly from the acoustic-scale value quoted by the reference fit.

Taking the recombination redshift to be $z_*=1089$, we define
$\theta_s\equiv r_s/D_*$, where
\begin{equation}
 D_*=\frac{c}{H_0}\int_0^{z_*}\frac{dz}{E(z)},
 \label{eq:acoustic-distance}
\end{equation}
and
\begin{equation}
 r_s(z)=\frac{c}{H_0}\int_z^{\infty}
 \frac{dz'}{E(z')\sqrt{3\left[1+
 \dfrac{3\omega_b^{\rm fid}}{4\omega_\gamma^{\rm fid}(1+z')}
 \right]}}.
 \label{eq:model-sound-horizon}
\end{equation}
We write $r_s\equiv r_s(z_*)$ when no argument is shown. Both quantities are evaluated using each model's expansion history.
Defining $\theta_{\rm eq}\equiv\theta(z_{\rm eq}^{\rm fid})$ at
$z_{\rm eq}^{\rm fid}\simeq3411.09$, the three positive unknowns
are obtained at fixed model inputs by solving
\begin{align}
 E^2(0)-1&=0
 &&\text{(budget equation)},
 \label{eq:fit-closure}\\
 \frac{h^2\Omega_{\rm Mon}A(\theta_{\rm eq})}
 {\omega_c^{\rm fid}}-1&=0
 &&\text{(DM matching at fiducial equality)},
 \label{eq:fit-cmb}\\
 \ln\!\left(\frac{r_s}{r_s^{\rm fid}}\right)
 +\ln\!\left(\frac{D_*^{\rm fid}}{D_*}\right)&=0
 &&\text{(acoustic angle)}.
 \label{eq:fit-acoustic}
\end{align}
Eq.~\eqref{eq:fit-cmb} matches the physical monopole density to the fiducial cold-dark-matter density at equality, while baryons remain a separate uncoupled component with $\Omega_b=\omega_b^{\rm fid}/h^2$. The final equation imposes $\theta_s^{\rm model}=\theta_s^{\rm fid}$.\footnote{After removing $a^{-3}$ dilution, the DM normalization changes by less than $0.36\%$ between equality and recombination and remains within $0.3\sigma$ of the Planck base-$\Lambda$CDM value $\omega_c=0.1200\pm0.0012$~\cite{Planck:2019nip}. The sampled axion energy fraction stays below $7.4\times10^{-4}$. These small background deviations support our approximation,
although they do not by themselves establish consistency with
the full CMB likelihood.}

More than one physical root can exist at the same input point. When a single surface must be displayed in our contour plots, we show the root with the smallest subsequently evaluated DESI $\chi^2_{\rm BAO}$. The appearance of a new root with a lower $\chi^2_{\rm BAO}$ accounts for jumps and discontinuities visible in some regions of the contour plots below. This is the pink  boundary visible in Fig.~\ref{fig:mu-ne-theta-f-chi2}.

We calibrate a fixed drag redshift, $z_d\simeq1059.90$, by requiring $r_s^{\rm fid}(z_d)=147.00\,\mathrm{Mpc}$. Each model then uses its own $r_d=r_s(z_d)$ in the BAO observables:
\begin{align}
 \frac{D_H(z)}{r_d}&=\frac{c}{H_0r_dE(z)},\\
 \frac{D_M(z)}{r_d}&=\frac{c}{H_0r_d}\int_0^z\frac{dz'}{E(z')},\\
 \frac{D_V(z)}{r_d}&=\left[z\left(\frac{D_M(z)}{r_d}\right)^2\frac{D_H(z)}{r_d}\right]^{1/3}.
\end{align}
At each DESI redshift we evaluate the observable specified by the published 13-component data vector and compute
\begin{equation}\label{eq:desi-chi2}
    \chi^2_{\rm BAO}
      = \bigl({\bf m}-{\bf d}\bigr)^{\!T}
        {\sf C}^{-1}
        \bigl({\bf m}-{\bf d}\bigr),
\end{equation}
where ${\bf d}$ is the DESI measurement vector, ${\bf m}$ the prediction of the constraint-solved background, and ${\sf C}$ the full covariance matrix from~\cite{DESI:2025zgx,DESI:2025zpo}. There are 13 observables. The analysis remains conditional on the fiducial calibration rather than constituting a full joint CMB+BAO likelihood.

Table~\ref{tab:wayne-ref} provides the DESI $\chi^2_{\rm BAO}$ values from the fits of an updated draft of Jhaveri et al.~\cite{Jhaveri:2026bla}. Our CMB-anchored procedure reproduces their fiducial $\Lambda$CDM value of $\chi^2_{\rm BAO}$ to good accuracy (see table from Fig.~\ref{fig:hz-weff-mu-sweep}), providing a consistency check on our CMB anchored BAO likelihood evaluation. The table therefore serves as a reliable numerical reference for the interested reader. However, the different numbers of adjustable cosmological parameters and the different fitting procedures must be kept in mind, so differences in these raw $\chi^2$ values alone should not be interpreted as a formal statistical preference between models.

\begin{table}[!t]
\centering
\begin{tabular}{lrcc}
\hline\hline
Model & $\chi^2_{\rm BAO}$ & $H_0$ [${\rm km\,s^{-1}\,Mpc^{-1}}$] & Extra parameters \\
\hline
Fiducial $\Lambda$CDM & 36.6 & 67.15 & --- \\
$\Lambda$CDM & 13.8 & 68.16 & --- \\
Early dark energy & 11.9 & 70.87 & 3 \\
$w_0w_a$CDM & 7.3 & 63.62 & 2 \\
\hline\hline
\end{tabular}
\caption{DESI DR2 BAO contributions reconstructed from Tables II and IV of the updated draft of Ref.~\cite{Jhaveri:2026bla}. The fiducial $\Lambda$CDM model is fitted to CMB data alone, while the remaining models are fitted to CMB+BAO without supernovae. The CMB combination includes ACT, SPT, and Planck. The final column counts additional cosmological parameters beyond the six of $\Lambda$CDM for the joint fits. These are raw BAO contributions, not total or reduced $\chi^2$ values.}
\label{tab:wayne-ref}
\end{table}

\subsection{Constraint surfaces and \texorpdfstring{$\chi^2_{\rm BAO}$}{chi-squared BAO}}

\begin{figure}[!htbp]
 \centering
 \includegraphics[width=\linewidth]{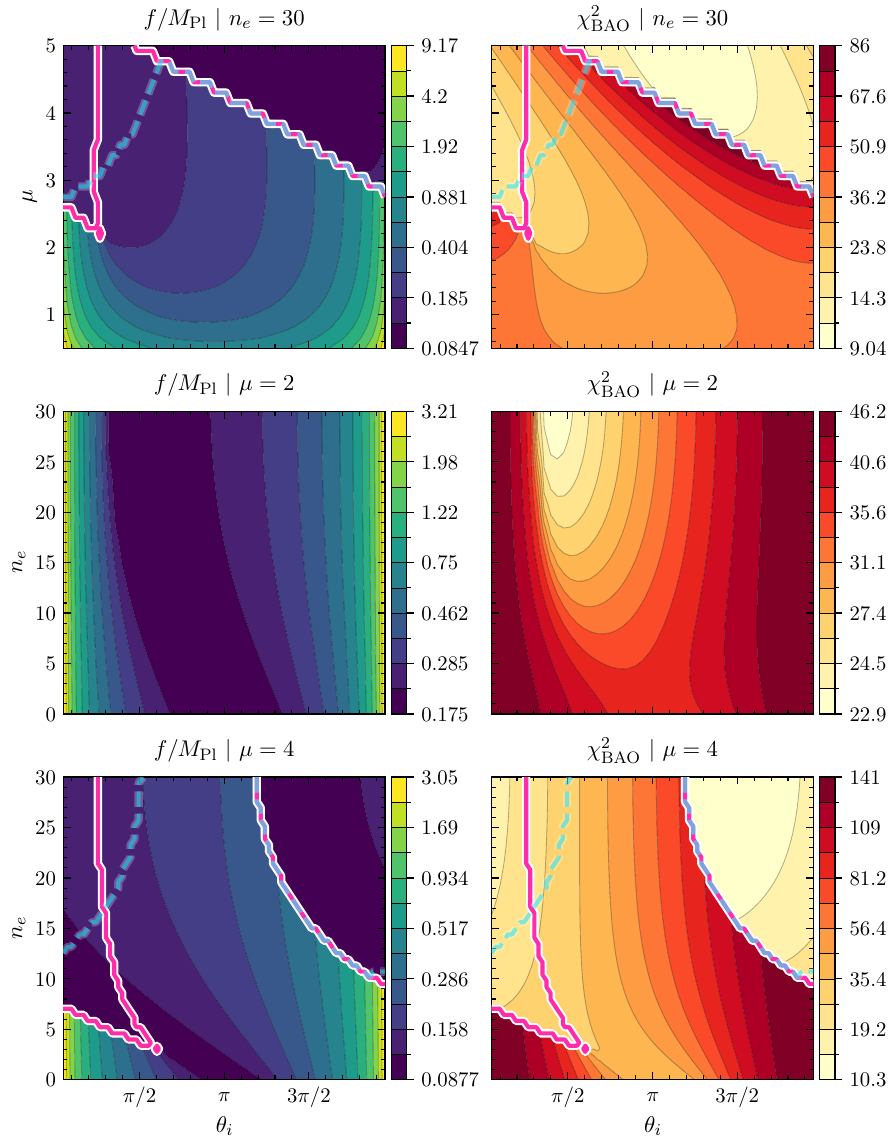}
 \caption{Constraint-solved $\fpl\equiv f_\varphi/\mpl$ (left) and predicted DESI DR2 $\chi^2_{\rm BAO}$ (right) at $\Omega_\Lambda=0.6$ and $\alpha_D=0.044$. \textbf{Top:} the $(\theta_i,\mu)$ plane at fixed $n_e=30$. \textbf{Middle and bottom:} the $(\theta_i,n_e)$ plane at $\mu=2$ and $4$, respectively. Regions bounded by bright pink lines mark multiple accepted roots; the root with the lowest $\chi^2_{\rm BAO}$ is displayed. Each panel has its own color scale; the contours are not confidence regions. The regions above the dashed cyan bands are disfavored by the fifth-force constraints from the CMB+BOSS fit discussed in Sec.~\ref{sec:astrophysconstraint}.}
 \label{fig:mu-ne-theta-f-chi2}
\end{figure}

In this section, we connect the field dynamics discussed above to the late-time BAO observables through the parameter scans in Fig.~\ref{fig:mu-ne-theta-f-chi2}. At each choice of model inputs, we determine $(\fpl,h,\Omega_{\rm Mon})$ jointly from Eqs.~\eqref{eq:fit-closure}--\eqref{eq:fit-acoustic} and subsequently evaluate $\chi^2_{\rm BAO}$. Several regions with smaller derived $\fpl$ exhibit better agreement with DESI. As we explain below, this correlation represents a general trend, while the detailed behavior can be more complicated and depends on the individual field histories.

The three rows of Fig.~\ref{fig:mu-ne-theta-f-chi2} make this pattern concrete. In the top row, increasing $\mu$ can allow a smaller $\fpl$ to satisfy the budget equations, since the normalized axion potential energy scales as $\mu^2\fpl^2$ at fixed field position. A smaller $\fpl$ in turn strengthens the monopole force through the explicit $\fpl^{-2}$ factor in the last term of Eq.~\eqref{eq:background-axion}, enhancing the monopole-driven motion and potentially producing a larger angular excursion during matter domination. The field can consequently cross potential barriers and even approach different cosine maxima by the time the monopole tilt has diluted (see Fig.~\ref{fig:axion-mechanism-tracking}). The broad picture that emerges is a preference in the DESI data for large axion field excursions and stronger dark matter--dark energy interactions within our model. However, $\fpl$ also enters through the expansion history, field velocity, and constraint-solved values of $h$ and $\Omega_{\rm Mon}$, so a complete picture requires examining individual trajectories; the preceding discussion should be taken as intuitive rather than exhaustive. The sharp drop in the upper-right region of this scan is associated with the appearance and selection of additional solution branches at higher $\mu$, as explained in more detail in Sec.~\ref{sec:multi-root}. The regions above the dashed cyan bands are disfavored by the fifth-force constraints from the CMB+BOSS fit discussed in Sec.~\ref{sec:astrophysconstraint}. 

The second and third rows of Fig.~\ref{fig:mu-ne-theta-f-chi2}, compared further in Sec.~\ref{sec:multi-root}, highlight a complementary feature: higher values of $n_e$ improve the DESI fit across the sampled range of $\fpl$, with the strongest improvement at lower $\fpl$. At a basic level, increasing $n_e$ at fixed $\Omega_{\rm Mon}$ and $\fpl$ strengthens the tilt toward larger $\theta$ for $\theta<2\pi n_e$, providing another handle on the matter-era excursion. However, the full picture is more involved because the $n_e^2$ term in $A(\theta)$ also increases the dyon mass, while the CMB normalization fixes $h^2\Omega_{\rm Mon}A(\theta_{\rm eq})$. Changes in the mass modulation are accordingly accompanied by a readjustment of the abundance and expansion rate, so the overall dependence on $n_e$ cannot be inferred from the tilt alone. The favorable regions at larger $n_e$ in Fig.~\ref{fig:mu-ne-theta-f-chi2} motivate our choice of $n_e=30$ for representative benchmarks. It is worth emphasizing that the key features of our model and its agreement with DESI do not rely on the particular choice $n_e=30$. The parameter $n_e$ simply provides a convenient additional degree of freedom.

\subsubsection{Multiple roots and branch switching}
\label{sec:multi-root}

\begin{figure}[!t]
    \centering
    \includegraphics[width=\textwidth]{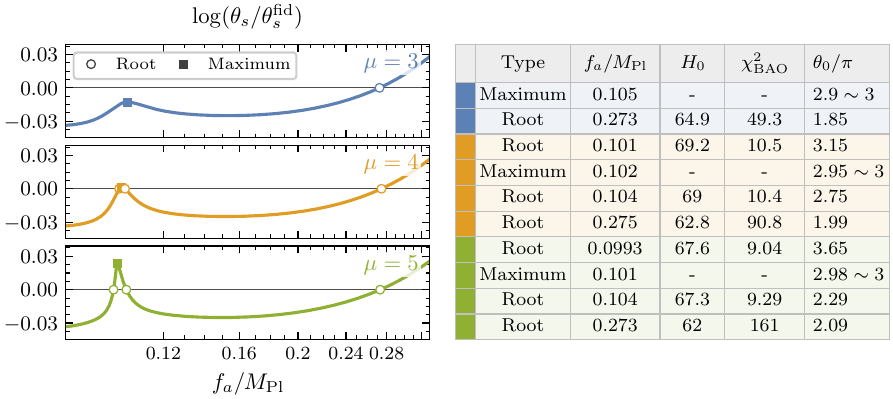}
    \caption{Emergence of additional constraint roots for $\theta_i=5\pi/4$, $n_e=30$, $\alpha_D=0.044$, and $\Omega_\Lambda=0.6$. \textbf{Left:} the acoustic residual $R$ of Eq.~\eqref{eq:constraint-acoustic-profile} for $\mu=3,4,5$, after imposing present-day closure and DM normalization at fiducial equality at each trial $\fpl$. Open circles mark solutions of all three constraints; filled squares mark local maxima of the residual, not additional constraint solutions. The low-$\fpl$ maximum lies below zero at $\mu=3$ and above zero at $\mu=4,5$, allowing two additional roots while the larger-$\fpl$ root persists. \textbf{Right:} the corresponding $\fpl$, $H_0$ in ${\rm km\,s^{-1}\,Mpc^{-1}}$, raw DESI DR2 $\chi^2_{\rm BAO}$, and present unwrapped field value $\theta_0/\pi$. Values of $H_0$ and $\chi^2_{\rm BAO}$ are reported only for accepted roots. The approximate integers beside the maximum entries indicate the field's proximity to $3\pi$; they do not imply an exactly stationary field.}
    \label{fig:root-resolution-combined}
\end{figure}

The sharp transitions marked by bright pink lines in Fig.~\ref{fig:mu-ne-theta-f-chi2} arise because the periodic axion potential allows the nonlinear constraint system to admit multiple backgrounds at identical model inputs. Their structure can be understood by solving closure and DM normalization at fiducial equality for $h(\fpl)$ and $\Omega_{\rm Mon}(\fpl)$ along a continuous branch and examining the remaining acoustic residual,
\begin{equation}
    R(\fpl;\mu)
    \equiv \ln\!\left[
        \frac{\theta_s(\fpl;\mu)}{\theta_s^{\rm fid}}
      \right]=\ln\!\left[
        \frac{r_s(\fpl;\mu)}{r_s^{\rm fid}}
      \right]
      + \ln\!\left[
        \frac{D_*^{\rm fid}}{D_*(\fpl;\mu)}
      \right],
    \label{eq:constraint-acoustic-profile}
\end{equation}
with all other parameters held fixed. At late times the cosine contribution to $D_*$ is substantial, so extrema of $D_*$ are expected to align with potential maxima at $\theta = (2k+1)\pi$ for integer $k$. This is confirmed in Fig.~\ref{fig:root-resolution-combined} and its table, where the peaks of $R(\fpl;\mu)$ appear near $\theta_0 \sim 3\pi$. As $\mu$ increases, these peaks grow sharper and can reach zero, creating a pair of roots at a fold where $R=\partial R/\partial\fpl=0$, while the pre-existing larger-$\fpl$ root persists.

For $\mu=3$, the low-$\fpl$ peak remains below zero and only the larger-$\fpl$ root is found. At $\mu=4$ and $5$, the peak becomes positive and two additional zero crossings appear near $\fpl\simeq0.10$, while the larger-$\fpl$ root remains near $\fpl\simeq0.27$. These solutions exhibit turning and barrier-crossing histories of the type illustrated in Fig.~\ref{fig:axion-mechanism-tracking}. At $\mu=4$, the two low-$\fpl$ roots, ordered by increasing $\fpl$, reach $\theta_0/\pi\simeq3.15$ and $2.75$, with positive and negative present-day velocities, respectively: one has passed the cosine maximum at $3\pi$ and is still advancing, whereas the other has turned back. The larger-$\fpl$ root instead reaches $\theta_0/\pi\simeq1.99$ and is still advancing.

For the illustrated $\mu=4$ solutions, the low-$\fpl$ roots retain a larger present axion fraction, including both kinetic and cosine-potential energy, and have higher $H_0$ and substantially lower $\chi^2_{\rm BAO}$ than the larger-$\fpl$ root. The two low-$\fpl$ solutions, ordered by increasing $\fpl$, have $H_0\simeq69.20$ and $69.03\,{\rm km\,s^{-1}\,Mpc^{-1}}$ and $\chi^2_{\rm BAO}\simeq10.48$ and $10.39$, compared with $H_0\simeq62.83\,{\rm km\,s^{-1}\,Mpc^{-1}}$ and $\chi^2_{\rm BAO}\simeq90.80$ for the larger-$\fpl$ solution. Although all three reproduce the fiducial acoustic angle $r_s/D_*=\theta_s^{\rm fid}$, this common ratio does not fix $D_M(z)/r_d$, $D_H(z)/r_d$, or $D_V(z)/r_d$ at individual BAO redshifts, which are evaluated using each model's $r_d$. The improvement is determined by these distance predictions and their covariance, rather than by $H_0$ or the present axion fraction alone.

The jumps into favorable regions consequently reflect switches between distinct constraint solutions rather than discontinuous evolution along a single branch. This is apparent when comparing the second and third rows of Fig.~\ref{fig:mu-ne-theta-f-chi2}: the fixed mass parameter changes from $\mu=2$ to $\mu=4$, allowing new low-$\fpl$ roots to become available in parts of the parameter plane. Where these roots provide a lower $\chi^2_{\rm BAO}$, selecting them produces the jumps in the displayed contours which is marked by the bright pink boundary lines. BAO data select among the accepted roots but do not determine their locations. The association of smaller $\fpl$ with better agreement therefore reflects the field and expansion histories discussed above.

\subsection{Low redshift behavior and DESI DR II BAO}\label{sec:desi-res}
% \hb{Refrences on instabiliy ($c_s^2<-1$: https://arxiv.org/abs/0709.1128 and https://arxiv.org/abs/astro-ph/0506663}

\begin{figure}[t]
    \centering
    \includegraphics[width=\linewidth]{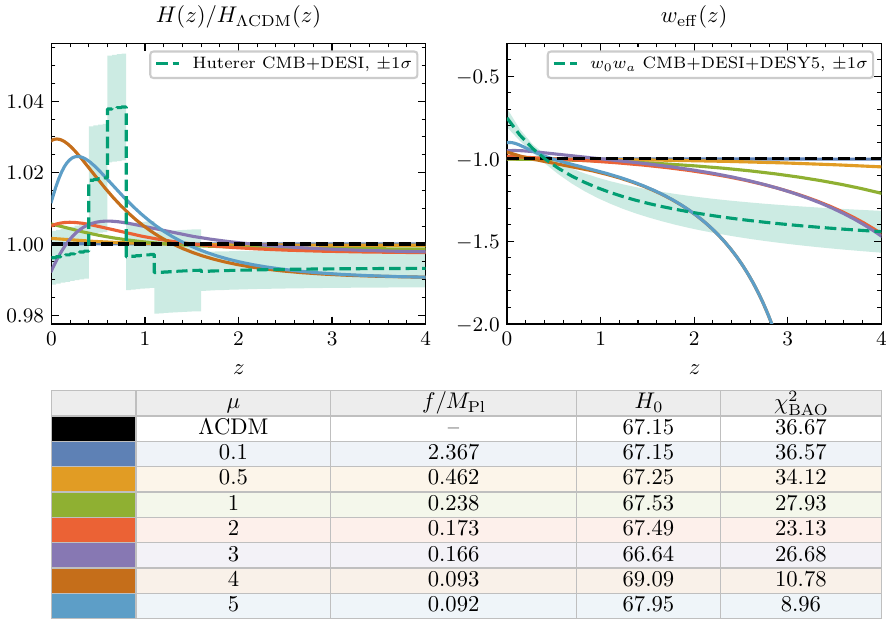}
    \caption{Expansion history and effective dark-energy equation of state for a sweep in $\mu=m_\varphi/H_0$, with $\Omega_\Lambda=0.65$, $\theta_i=3\pi/4$, $\alpha_D=0.044$, and $n_e=30$. \textbf{Left:} $H(z)$ relative to the fiducial $\Lambda$CDM expansion rate. The green dashed curve and shaded band show the CMB+DESI modified-$H$ reconstruction and its $1\sigma$ interval from Bansal and Huterer~\cite{Bansal:2025ipo}. \textbf{Right:} the corresponding $w_{\rm eff}(z)$, compared with the CMB+DESI best-fit $w_0w_a$ history and its $1\sigma$ interval from Ref.~\cite{DESI:2025zgx}. Black dashed lines denote the fiducial $\Lambda$CDM predictions. For each model, $(\fpl,h,\Omega_{\rm Mon})$ are determined jointly by the three background constraints. The color-matched table reports the derived parameters and the raw $\chi^2_{\rm BAO}$ for the 13 DESI DR2 observables, with $H_0$ in ${\rm km\,s^{-1}\,Mpc^{-1}}$. The comparison bands are illustrative and are not included as additional data in $\chi^2_{\rm BAO}$.}
    \label{fig:hz-weff-mu-sweep}
\end{figure}

The DESI BAO data can be interpreted as evidence for an evolving dark-energy density at low redshift~\cite{DESI:2025zgx, DES:2025sig}. This evolution can be characterized in two ways: directly through the expansion rate $H(z)$, or through an effective equation of state $w_{\rm eff}(z)$, where values below $-1$ signal apparent phantom behavior. In this section, we show that the axion--monopole interaction naturally reproduces both signatures and provides a competitive fit to the DESI data.

\subsubsection{Expansion-rate comparison}

In the cosmological model-building literature, the impact of DESI data is often illustrated by adopting a smooth equation-of-state ansatz and looking for phantom behavior. A less restrictive diagnostic is the modified-$H$ reconstruction of Bansal and Huterer~\cite{Bansal:2025ipo}, in which $H(z)$ is allowed to vary independently in the DESI redshift bins and is constrained by DESI DR2 BAO and a compressed CMB likelihood. This approach avoids assuming a particular smooth form for the dark-energy equation of state, so we first compare our expansion histories with their reconstruction before turning to phantom behavior. The authors emphasize that this construction is flexible rather than strictly model-independent, and we therefore use it only as a visual comparison. Quantitative statements in the table below Fig.~\ref{fig:hz-weff-mu-sweep} use the 13-component DESI BAO statistic from Eq.~\eqref{eq:desi-chi2}.

The left panel of Fig.~\ref{fig:hz-weff-mu-sweep} shows that the axion--dark monopole interaction produces a smooth, percent-level deformation of $H(z)$. Along the $\mu$ sweep, the lowest raw $\chi^2_{\rm BAO}$ values occur for the larger displayed masses, which correspond to smaller derived $\fpl$. These solutions also produce a more pronounced low-redshift enhancement of $H(z)$, although they do not fully reproduce the sharp peak below $z\sim1$ in the Bansal--Huterer reconstruction. Notably, the lowest-$\chi^2_{\rm BAO}$ points in this sweep lie on new solution branches that appear at higher $\mu$. Comparing the table in Fig.~\ref{fig:hz-weff-mu-sweep} with Table~\ref{tab:wayne-ref}, our best benchmarks have lower raw BAO $\chi^2$ values than the $\Lambda$CDM and EDE reference fits and are competitive with $w_0w_a$CDM. Relative to the latter, they yield substantially higher values of $H_0$ at comparable $\chi^2_{\rm BAO}$. This remains an indicative comparison, however, because the numbers of adjustable parameters and the fitting procedures differ. A full joint likelihood analysis is needed to assess the relative statistical support for these models.

\subsubsection{Effective equation of state and phantom behavior}\label{sec:phantom_behavior}

As discussed above, interpreting DESI constraints in terms of an effective dark-energy equation of state requires specifying a separately conserved CDM component. To compare with this description, we define the fiducial CDM density, anchored at fiducial equality by the same condition used in Eq.~\eqref{eq:fit-cmb},
\begin{equation}
\rho_c^{\rm fid}(x)
\equiv \rho_{\rm crit}\Omega_{\rm Mon}A(\theta_{\rm eq})e^{-3x}
=\rho_{\rm crit}\frac{\omega_c^{\rm fid}}{h^2}e^{-3x}.
\label{eq:cmb-normalized-cdm}
\end{equation}

The remaining energy density and pressure are assigned to an effective dark-energy component,
\begin{align}
\rho_{\rm DE}^{\rm eff}(x)
&=\rho_\Lambda+\rho_\theta(x)+\rho_{\rm Mon}(x)-\rho_c^{\rm fid}(x)\nonumber\\
&=\rho_\Lambda+\rho_\theta(x)+\rho_{\rm Mon}(x)
\left[1-\frac{A(\theta_{\rm eq})}{A(\theta)}\right],
\label{eq:rho-de-effective}\\
p_{\rm DE}^{\rm eff}(x)&=-\rho_\Lambda+p_\theta(x).
\end{align}
The interaction correction therefore vanishes at fiducial equality. We also define
\begin{equation}
w_{\theta\Lambda}(x)
\equiv \frac{p_\theta(x)-\rho_\Lambda}
{\rho_\theta(x)+\rho_\Lambda},
\end{equation}
for which $\rho_\theta+\rho_\Lambda>0$, and the canonical axion kinetic term ensures $w_{\theta\Lambda}\geq-1$. The inferred equation of state is then~\cite{Das:2005yj, Khoury:2025txd}
\begin{equation}
w_{\rm eff}(x)
\equiv\frac{p_{\rm DE}^{\rm eff}(x)}{\rho_{\rm DE}^{\rm eff}(x)}
=\frac{w_{\theta\Lambda}(x)}{
1+\dfrac{\rho_{\rm Mon}(x)}{\rho_\theta(x)+\rho_\Lambda}
\left[1-\dfrac{A(\theta_{\rm eq})}{A(\theta)}\right]}
=\frac{w_{\theta\Lambda}(x)}{
1+\dfrac{\Delta\rho_{\rm Mon}(x)}{\rho_\theta(x)+\rho_\Lambda}},
\label{eq:w_eff}
\end{equation}
where in the last line we took $\Delta\rho_{\rm Mon}(x)
\equiv \rho_{\rm Mon}(x)
\left[1-A(\theta_{\rm eq})/A(\theta)\right]$.
For positive $\rho_{\rm DE}^{\rm eff}$, we find that effective phantom behavior occurs when
\begin{align}
w_{\rm eff}<-1
&\quad\Longleftrightarrow\quad
\rho_{\rm DE}^{\rm eff}+p_{\rm DE}^{\rm eff}<0 \nonumber\\
&\quad\Longleftrightarrow\quad
\Delta\rho_{\rm Mon}<-\bigl(\rho_\theta+p_\theta\bigr)
=-f_\varphi^2H^2(\theta')^2.
\label{eq:phantom-condition}
\end{align}

The two classes of late-time trajectories in Fig.~\ref{fig:axion-mechanism-tracking}, one crossing a cosine barrier and the other turning around, produce distinct phantom phenomenology. Throughout the field range $\theta<2\pi n_e$, $dA/d\theta<0$, so it is the direction of motion, not barrier crossing per se, that controls the sign of the monopole mass change. A trajectory advancing toward larger $\theta$ accumulates a growing deficit in $A$ relative to its equality value, while one that turns around begins to restore $A$. This connects directly to Eq.~\eqref{eq:phantom-condition}: $A(\theta)<A(\theta_{\rm eq})$ is necessary but not sufficient for phantom crossing, as the density deficit must also exceed the positive axion kinetic contribution. A barrier-crossing trajectory therefore accumulates a larger deficit and can sustain the phantom phase, whereas a turnaround reduces the deficit while retaining comparable kinetic energy. Whether the phantom phase persists depends on the interplay between these two contributions.

The right panel of Fig.~\ref{fig:hz-weff-mu-sweep} shows the effective equation of state defined in Eq.~\eqref{eq:w_eff} across a sweep in $\mu$ at fixed remaining parameters. The higher-mass, low-$\fpl$ benchmarks show more pronounced phantom behavior at intermediate redshifts. We superpose the CMB+DESI+DESY5 $w_0w_a$CDM fit for visual comparison. The two lowest-$\chi^2_{\rm BAO}$ points in this sweep correspond to the heaviest masses and both arise from new branches at lower $\fpl$. It is also worth noting that the $\mu=5$ benchmark exhibits the largest (least negative) $w_{\rm eff}(z=0)$ among the displayed models.

\subsection{Birefringence}\label{sec:birefringence}

The axion trajectories discussed above also generate cosmic birefringence if the field couples to visible-sector photons through
\begin{equation}
 \mathcal{L}_{\theta\gamma}
 =-\frac{\alpha_{\rm EM}c_\gamma}{8\pi}\,
   \theta F_{\mu\nu}\widetilde F^{\mu\nu},
 \label{eq:birefringence-coupling}
\end{equation}
where $c_\gamma$ is the photon anomaly coefficient, $F_{\mu\nu}$ the electromagnetic field strength, and $\widetilde F^{\mu\nu}$ its dual. We distinguish $\alpha_{\rm EM}$ from $\alpha_D$ and neglect photon backreaction on the background solutions of Sec.~\ref{sec:desi-method}

For a homogeneous axion in a spatially flat FRW background, we introduce conformal time $d\eta=dt/a$ and decompose the transverse photon field into circular polarizations, $A_\pm=(A_x\mp iA_y)/\sqrt{2}$. In Coulomb gauge, their equations of motion are~\cite{Gasparotto:2022uqo}
\begin{equation}
 \left(\partial_\eta^2+\omega_\pm^2\right)A_\pm=0,
 \qquad
 \omega_\pm^2=k^2\mp k\frac{\alpha_{\rm EM}c_\gamma}{2\pi}
                    \frac{d\theta}{d\eta},
 \label{eq:birefringence-gauge-eom}
\end{equation}
where $k$ is the comoving wavenumber. For small couplings the two helicities acquire frequencies $\omega_\pm\simeq k\mp\alpha_{\rm EM}c_\gamma(d\theta/d\eta)/(4\pi)$, and their relative phase accumulation rotates the plane of linear polarization. Taking positive $\beta$ to denote clockwise rotation, photons emitted at $\eta_{\rm em}$ and observed at $\eta_0$ acquire
\begin{equation}
 \beta=\frac{1}{2}\int_{\eta_{\rm em}}^{\eta_0}
                   (\omega_--\omega_+)\,d\eta
 =\frac{\alpha_{\rm EM}c_\gamma}{4\pi}
   \left[\theta(\eta_0)-\theta(\eta_{\rm em})\right].
 \label{eq:birefringence-phase-rotation}
\end{equation}
The rotation therefore depends on the net axion displacement between emission and observation, rather than on the total distance traveled by the field.

For CMB photons, emission corresponds to last scattering, which occurs over a finite interval rather than at a single instant. Following Ref.~\cite{Fujita:2020ecn}, we account for this by weighting the axion trajectory with the visibility function $\mathcal{V}(T)$, which describes the probability density for a CMB photon to last scatter at temperature $T$.
\begin{equation}
 \begin{aligned}
 \overline\theta_{\rm LSS}
 &=\int dT\,\mathcal{V}(T)\,\theta(x(T)),\\
 \mathcal{V}(T)
 &\simeq\frac{1}{\sqrt{2\pi}\sigma_T}
 \exp\!\left[-\frac{(T-T_{\rm LSS})^2}{2\sigma_T^2}\right],
 \end{aligned}
 \label{eq:birefringence-average}
\end{equation}
with $T_{\rm LSS}=2941\,\mathrm{K}$, $\sigma_T=248\,\mathrm{K}$, and $x(T)=\ln(T_0/T)$, where $T_0$ is the present CMB temperature. If the field changes little during last scattering, this average is well approximated by its value at $T=T_{\rm LSS}$. Rapid oscillations can instead lead to cancellations that reduce the last-scattering contribution to the net rotation.

The visibility-averaged rotation is
\begin{equation}
 \beta=\frac{\alpha_{\rm EM}c_\gamma}{4\pi}
       \Delta\theta_{\rm LSS},
 \qquad
 \Delta\theta_{\rm LSS}\equiv\theta_0-\overline\theta_{\rm LSS},
 \label{eq:birefringence-angle}
\end{equation}
where $\beta$ is in radians. We adopt $\beta=0.30^\circ$ as our benchmark rotation. For comparison, the nearly full-sky Planck PR4 analysis reports $\beta=0.30^\circ\pm0.11^\circ$~\cite{Diego-Palazuelos:2022dsq}, the joint WMAP and Planck analysis gives $\beta=0.342^{\circ\,+0.094^\circ}_{\phantom{\circ}\,-0.091^\circ}$~\cite{Eskilt:2022cff}, ACT DR6 finds $\beta=0.215^\circ\pm0.074^\circ$~\cite{Diego-Palazuelos:2025dmh}, and the recent joint ACT and Planck analysis yields $\beta=0.277^\circ\pm0.057^\circ$~\cite{Eskilt:2026imm}, all at 68\% confidence. The photon anomaly coefficient required to produce the benchmark $\beta=0.30^\circ$ is then
\begin{equation}
 c_\gamma^{\rm target}\equiv |c_\gamma|
 =\frac{4\pi|\beta|}{\alpha_{\rm EM}|\Delta\theta_{\rm LSS}|}
 =\frac{9.0142}{|\Delta\theta_{\rm LSS}|}
   \left(\frac{|\beta|}{0.30^\circ}\right)
   \left(\frac{1/137}{\alpha_{\rm EM}}\right).
 \label{eq:birefringence-anomaly}
\end{equation}
The explicit $f_\varphi$ dependence cancels, but $\fpl$ still affects the axion trajectory. For a fixed rotation, a small net displacement demands a large photon anomaly coefficient. The monopole interaction can instead enhance the displacement, reducing the coefficient required from the UV completion.

\begin{figure}[t]
 \centering
 \includegraphics[width=\linewidth]{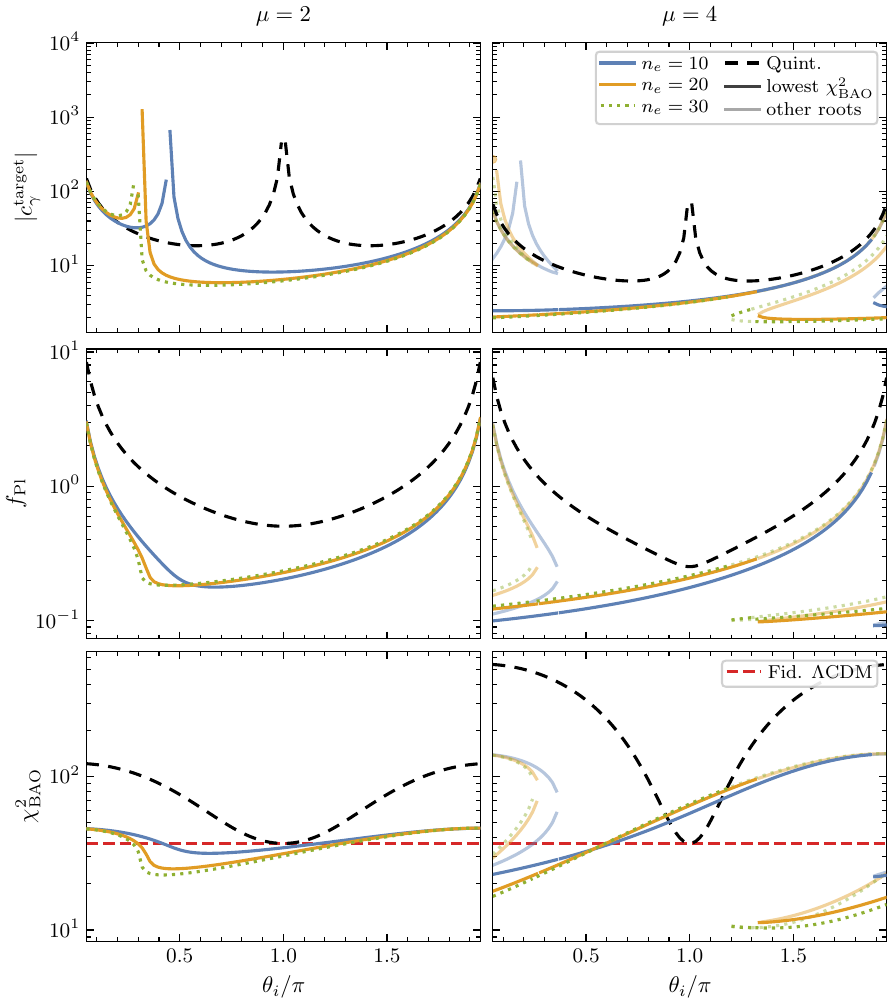}
 \caption{\textbf{Top:} photon anomaly coefficient $|c_\gamma|$ required for $\beta=0.30^\circ$ with $\alpha_{\rm EM}=1/137$, \textbf{Middle:} derived $\fpl$, and \textbf{Bottom:} raw $\chi^2_{\rm BAO}$ for the 13 DESI DR2 observables, all as functions of $\theta_i$ at $\mu=2$ (left) and $\mu=4$ (right). Colored curves show the interacting model with $\Omega_\Lambda=0.6$, $\alpha_D=0.044$, $n_e=10,20$, and $n_e=30$ dotted. Black dashed curves show uncoupled quintessence with $\alpha_D=0$ and $\Omega_\Lambda=0$. At each $\theta_i$, $(\fpl,h,\Omega_{\rm Mon})$ are fixed by the three background constraints. The quintessence curves correspond to the unique recovered root at each angle, not a globally optimized DESI fit. All recovered roots are shown; solid colored curves mark the lowest-$\chi^2_{\rm BAO}$ root and faded curves show alternatives.}
 \label{fig:birefringence-theta-i}
\end{figure}

Fig.~\ref{fig:birefringence-theta-i} compares the interacting model with uncoupled axion quintessence. In both cases $(\fpl,h,\Omega_{\rm Mon})$ are solved jointly from Eqs.~\eqref{eq:fit-closure}--\eqref{eq:fit-acoustic} using the same fiducial calibration. The quintessence reference sets both $\alpha_D$ and $\Omega_\Lambda$ to zero, so the comparison changes the dark-energy composition as well as the coupling. Between the two panels, the interacting model permits smaller coefficients, with the largest contrast near initial angles where quintessence produces little net displacement.

This difference can be traced to the forces driving the axion. In uncoupled cosine quintessence, the driving term is proportional to $m_\varphi^2\sin\theta$. For $m_\varphi$ of order $H_0$, substantial motion is delayed until late times and further suppressed near the potential hilltop. The uncoupled curve is consequently symmetric about $\theta_i=\pi$, where an exact hilltop initial condition with zero velocity yields no rotation. In our model, the monopole-induced tilt discussed in Sec.~\ref{sec:tracking} remains nonzero at the hilltop. Because its strength relative to Hubble friction is approximately sustained during matter domination, the field can undergo appreciable rolling long before the intrinsic cosine potential becomes dynamically important. The tilt therefore breaks the exact hilltop stationarity of the uncoupled model, but does not guarantee a nonzero net displacement for every initial angle.

Indeed, the interacting curves at $\mu=2$ still exhibit strong sensitivity to $\theta_i$ near points where $\theta_0-\overline\theta_{\rm LSS}$ vanishes and $c_\gamma^{\rm target}$ diverges. Such a cancellation can result from a trajectory that rolls and then turns back toward its last-scattering value, rather than from a field that remains stationary. 

The absence of these asymptotes from the selected $\mu=4$ curves has a different origin. As discussed in Sec.~\ref{sec:multi-root}, higher $\mu$ opens additional low-$\fpl$ branches that are preferred by the DESI data. These same branches retain sizable net displacements where the larger-$\fpl$ solutions approach a cancellation, so selecting the lowest-$\chi^2_{\rm BAO}$ root simultaneously avoids the divergent $c_\gamma^{\rm target}$ regions. This is particularly evident for $n_e=20$ and $30$, where the selected curves remain at coefficients of a few across the sampled angular range. The divergent branches persist among the alternative solutions, and switching between roots can still produce finite jumps.

\subsubsection{Implications for model building}
\label{sec:birefringence-uv}

The smaller photon anomaly coefficients have a useful model-building implication. Large effective coefficients can in principle be generated, but their origin must be specified in a UV completion and may face its own challenges. For example, enhancing the anomaly through large electric charges or a large multiplicity of charged particles can encounter perturbativity constraints, as additional hypercharged particles above their mass thresholds accelerate the running of the $U(1)_Y$ coupling and can lower its Landau-pole scale below the intended UV cutoff~\cite{Agrawal:2017cmd}.

Our model reduces the need for such enhancement by increasing the axion displacement through the same dark-sector interaction that shapes the BAO predictions. The low-$\fpl$ branches identified above accommodate the benchmark rotation with modest coefficients over broad ranges of the initial angle, easing the demands on the photon-coupling sector relative to uncoupled quintessence.

\subsection{Astrophysical constraints on dark magnetic monopoles} \label{sec:astrophysconstraint}
In this section, we consider two types of monopole interactions, namely dark electromagnetic interactions mediated by the dark photon and the axion-mediated fifth force arising from the field dependence of the dyon mass.

First, we consider dark electromagnetic interactions. A dark magnetic monopole of mass $M$ interacting via an (effectively) massless dark photon has Rutherford/Coulomb scattering, just like electrically charged DM. Since the magnetic charge of monopoles will dominate their interactions, we need to replace $\alpha_D$ by $\alpha_M$ in the momentum-transfer cross section,
\begin{equation}
    \sigma_{\rm tr}\sim \frac{4\pi\alpha_M^2}{M^2v^4}\ln\Lambda_c,
\end{equation}
where $v$ is the dark matter velocity in dark matter halos, $\Lambda_c$ is given by the ratio of the largest to the smallest impact parameter, and the magnetic fine-structure constant can be written as $\alpha_M=1/\alpha_D$. Since we consider superheavy magnetic monopoles $M\sim 10^{6}$ GeV, the self-interacting cross-section is tiny ($\sigma_{\rm tr}/M\sim10^{-5}{\rm cm^2/g}$ for $v\sim 10^{-3}$) and the traditional constraints on self-interacting dark matter will not apply.

Next, we consider plasma screening and collective effects of the same dark electromagnetic interaction. Since both monopoles and anti-monopoles will be present in dark matter halos, their dark forces tend to screen each other in the long range. For a thermal/virialized monopole plasma with number density $n$ and mass density $\rho$, the Debye length is parametrically the same as that of electric plasmas
\begin{equation}
    \lambda_D \sim \sqrt{\frac{T}{4\pi \alpha_M n}}\approx \frac{M v}{\sqrt{4\pi \alpha_M \rho}}.
\end{equation}
Taking the cosmic average density and a galactic virial velocity $v\approx 10^{-3}$ as a benchmark, we find
\begin{equation}
    \lambda_D\approx 1.66\times 10^{-3}{\rm pc} \left(\frac{M}{10^{10} \rm GeV}\right) \left(\frac{10}{\alpha_M}\right)^{1/2}.
\end{equation}
Note that the velocity can be smaller in dwarf galaxies, which lead to an even smaller Debye radius. Therefore, the magnetic interactions between monopoles will not affect large-scale structures as gravity for the relevant mass ranges we consider here. An interesting possibility is that the dark monopoles might generate dark electromagnetic fields with large coherent lengths, which leads to collective effects on their scattering even when the dark charged particles are superheavy \cite{Cruz:2022otv,DeRocco:2022rze,DeRocco:2024ifs}. However, the charged particles considered in this work are dyons with both dark magnetic and electric charges, which may exhibit different dynamics with dark EM fields, which we leave for future exploration.

Lastly, we consider the axion-mediated fifth force. The long-range self-interactions of dark matter also enhance the matter power spectrum~\cite{Bottaro:2024pcb}, which is another testable feature of this class of models.

We denote the dimensionless fifth-force coupling by $\beta_{\rm fifth}$ to distinguish it from the birefringence angle $\beta$. In terms of a canonically normalized field $\varphi$, Ref.~\cite{Bottaro:2024pcb} constrains a dark matter mass of the form
\begin{equation}
    m_\chi(\varphi) = m_\chi \left(1 - \sqrt{\frac{\beta_{\rm fifth}}{2}} \frac{\varphi}{\mpl}\right).
\end{equation}
We use their mass-dependent 95\% confidence-level upper bound $\beta_{{\rm fifth},95}(m_\varphi/H_0)$ from the CMB+BOSS analysis, combining Planck CMB data with the BOSS full-shape galaxy power spectrum and earlier BAO measurements. Comparing this to our assumptions, we rewrite their parameter $\beta_{\rm fifth}$ as
\begin{equation}
    \beta_{\rm fifth} = 2 \left(\frac{\gamma \alpha_D^2 n_e \mpl}{2 \pi f_\varphi} \frac{1}{1+\gamma \alpha_D^2n_e^2/2}\right)^2.
\end{equation}
We translate the published CMB+BOSS upper-limit curve into an approximate lower bound on $f_\varphi$ at each grid point. This gives
\begin{equation}
    \frac{f_{\varphi,\min}}{\mpl} = \left|\frac{\gamma \alpha_D^2 n_e}{2\pi\left(1+\gamma\alpha_D^2n_e^2/2\right)}\right|\sqrt{\frac{2}{\beta_{{\rm fifth},95}(\mu)}}.
\end{equation}
We evaluate this bound with $\gamma=1$ and the values of $\mu$, $n_e$, and $\alpha_D$ at each grid point. Regions above the cyan boundaries in Fig.~\ref{fig:mu-ne-theta-f-chi2} delineate solutions with $f_\varphi<f_{\varphi,\min}$ and are therefore disfavored under this approximate translation. Where multiple constraint roots exist, other roots at the same grid point may still satisfy the bound. Ref.~\cite{Bottaro:2024pcb} also presents stronger projected constraints obtained by adding full-shape galaxy-clustering forecasts for DESI, Euclid, or MegaMapper to CMB+BOSS. Applying these projected limits through the same translation would strengthen the lower bound on $f_\varphi$ and enlarge the disfavored region. 

For our benchmark parameters $\alpha_D=0.044$, $\gamma=1$, and $n_e=30$, the approximate CMB+BOSS limits of Ref.~\cite{Bottaro:2024pcb} give $\beta_{\rm fifth}\lesssim5\times10^{-3}$ for $m_\varphi<H_0$ and $\beta_{\rm fifth}\lesssim2\times10^{-3}$ at $m_\varphi=4H_0$, translating into $f_{\varphi,\min}\simeq0.10\,\mpl$ and $0.16\,\mpl$, respectively.

\subsection{Tension with the Axion Weak Gravity Conjecture} \label{sec:AWGC}

A possible theoretical constraint on our scenario arises from the Axion Weak Gravity Conjecture (AWGC), which asserts that for an axion of decay constant $f$, there should exist a charge-$q$ instanton with action $S$ such that $fS \lesssim |q| \mpl$~\cite{Arkani-Hamed:2006emk}. The analogy with the standard Weak Gravity Conjecture (WGC) was initially not precise enough to put a definite coefficient in this bound, but recently~\cite{DiUbaldo:2026rly, Maldacena:2026jqd} gave a precise argument in favor of a quantitative bound originally proposed in~\cite{Montero:2015ofa}:
\begin{equation} \label{eq:AWGC}
    fS \leq |q| \frac{\pi}{2}\sqrt{\frac{3}{2}} \mpl.
\end{equation}
Supersymmetric examples in string theory obey a slightly tighter bound~\cite{Etheredge:2026rio}. This can be a strong constraint on models of axion dark energy, as discussed in~\cite{Rudelius:2022gyu} for Early Dark Energy and recently by~\cite{Shiu:2026edl} for a dark energy axion with a periodic potential. The considerations they discuss apply almost directly to our scenario, although the details of our model are somewhat different so quantitative details may differ slightly.

As discussed in Sec.~\ref{sec: modelsetup}, we are interested in $\alpha_D \approx 0.044$, for which $SU(2)'$ instantons have $S_D = 2\pi/\alpha_D \approx 140$. In this case, the axion WGC requires
\begin{equation}
    f_\varphi = f/|q| \leq 1.3 \times 10^{-2} \mpl = 3.2 \times 10^{16}\,\mathrm{GeV}.
\end{equation}
This is well below the values of $f_\varphi$ that provide good fits in Fig.~\ref{fig:mu-ne-theta-f-chi2}. Furthermore, it is in direct conflict with the bound discussed in Sec.~\ref{sec:astrophysconstraint} from the axion-induced long-range force between dark matter particles, which requires $f_\varphi$ to be an order of magnitude larger.

Several years ago there was a vigorous effort to understand the constraints imposed by the axion WGC on models of axion inflation, which identified a number of potential loopholes; see, e.g.,~\cite{delaFuente:2014aca,Rudelius:2015xta,Montero:2015ofa,Brown:2015iha,Bachlechner:2015qja,Hebecker:2015rya,Brown:2015lia,Heidenreich:2015wga,Heidenreich:2019bjd}. The easiest to apply to our scenario (as in the similar discussion in~\cite{Rudelius:2022gyu}) is the ``extra-instanton loophole'': if the AWGC bound in Eq.~\eqref{eq:AWGC} is obeyed by a higher-charge instanton (unrelated to $SU(2)'$ dynamics), while $SU(2)'$ contributes $q = 1$ instantons that violate the AWGC, the AWGC-obeying instantons may not contribute significantly to the potential. Such a scenario seems difficult to realize in UV completions, but at least in principle gives a logical possibility to make our model compatible with the AWGC. Alternatively, our argument for $\alpha_D \approx 0.044$ was really an argument about the size of the dark energy potential. If this size is suppressed relative to the instanton estimate (e.g., using the $\mathbb{Z}_N$ mechanism of~\cite{Hook:2018jle}), we could consider smaller values of $S$ for a given $f_\varphi$. Apart from being potentially baroque, this would raise new puzzles (e.g., why is only one of the $N$ sectors populated in cosmology?), but again serves as a logical loophole. We leave more thorough consideration of how to find a realization of our mechanism compatible with the AWGC for future work.

\section{Conclusion}

In this work, we have presented a natural framework for interacting dark matter and dark energy based on axion--monopole dynamics. Dark matter is composed of dark magnetic monopoles, while an ultralight axion provides the dynamical dark-energy component. Through the Witten effect, the axion changes the electric charge and hence the mass of the monopoles, generating the interaction between the axion dark energy and monopole dark matter. In our setup, the monopole spectrum forms a monodromy tower of dyonic states. Summing over this tower can exponentially suppress the radiative potential, avoiding the naturalness problem that generically applies to DM--DE interacting models which generate a large Coleman-Weinberg potential for the dark energy field.

In the minimal realization, the interaction from the Witten effect is relatively small, at the ${\cal O}(10^{-3})$ level. We have shown, however, that the effective coupling can be enhanced to the percent level if higher-charge dyonic states are populated asymmetrically in the early Universe. The resulting monopole-induced potential naturally tracks the Hubble scale during matter domination and can drive an appreciable axion evolution. This dynamics produces a modified late-time expansion history and can lead to an apparent phantom equation of state. The same axion evolution can also generate cosmic birefringence while requiring only modest photon anomaly coefficients, connecting two potentially observable consequences of the dark-sector interaction. 
\acknowledgments

We thank Prateek Agrawal, JiJi Fan, Marios Galanis, Wayne Hu, and Rayne Liu for useful conversations. We acknowledge the use of ChatGPT and Claude in preparing portions of numerical code and cross-checking calculations. All AI-generated code was benchmarked against independently written human code. HB is supported by the National Science Foundation MPS-Ascend Postdoctoral Research Fellowship under Award No.~2503442. MR is supported in part by the DOE Grant DE-SC0013607. LTW is supported by the DOE Grant DE-SC0009924. HX is supported by the DOE Grant DE-SC0026297. This work was performed in part at the Aspen Center for Physics, which is supported by National Science Foundation Grant PHY-2210452.
\appendix

\section{Wave dark matter--dark energy interactions}
\label{app:waveDMDE}

The general naturalness argument that we gave in Sec.~\ref{subsec:naturalnessproblem} left open the possibility that $\chi$ can be wave dark matter (with a mass below the eV scale) while remaining compatible with small radiative corrections to the $\varphi$ potential. In this appendix, we will further argue that it is not straightforward to build a model along these lines.

We will consider $\chi$ and $\varphi$ both to be axion-like fields, with periodicity $\chi \cong \chi + 2\pi f_\chi$ and $\varphi \cong \varphi + 2\pi f_\varphi$. The simplest type of model to consider has two periodic potentials, with mass and kinetic mixing:
\begin{equation}
    {\cal L}_0 = - \frac{1}{2} (\partial \varphi)^2 - \frac{1}{2} (\partial \chi)^2 - \kappa\, \partial \varphi \cdot \partial \chi + \Lambda_1^4 \cos(k\,\chi/f_\chi) + \Lambda_2^4 \cos(p\, \chi/f_\chi + q\, \varphi/f_\varphi), \quad k, p, q \in \mathbb{Z}.
\end{equation}
We assume that $k^2 \Lambda_1 \gg p^2 \Lambda_2$, so that the dark matter field $\chi$ primarily gets its mass from the first potential term. There is no loss of generality in making this term depend only on $\chi$, since we could always perform a $\mathrm{GL}(2,\mathbb{Z})$ transformation on ($\chi/f_\chi, \varphi/f_\varphi)$ to achieve this form. There is also no loss of generality in not adding constant phase offsets in the arguments of the cosine, since we can shift our definition of $\chi$ and $\varphi$ by constants to absorb such a phase.

We are interested in the cosmological era where $\chi$ behaves as dark matter, meaning that it oscillates about $\chi = 0$ in the quadratic part of its potential: 
\begin{equation}
    \chi(t) \approx \chi_0(t) \sin(m_\chi (t - t_0)).
\end{equation}
We can eliminate the kinetic mixing (at the cost of losing a basis with simple periodicity properties) by the field redefinition
\begin{equation}
    \varphi = {\tilde \varphi} - \kappa \chi. 
\end{equation}
The effect of the dark matter field $\chi$ on the ${\tilde \varphi}$ potential is determined by the average over $\chi$ oscillations:
\begin{align}
    \frac{\partial V}{\partial {\tilde \varphi}} = q \frac{\Lambda_2^4}{f_\varphi}\langle \sin(p\,\chi/f_\chi + q\,\varphi/f_\varphi)\rangle &= q \frac{\Lambda_2^4}{f_\varphi} \sin(q\, {\tilde \varphi}/f_\varphi) \langle \cos(p\,\chi/f_\chi - \kappa\, q \chi/f_\varphi)\rangle \nonumber \\ &\approx q \frac{\Lambda_2^4}{f_\varphi} \sin(q\, {\tilde \varphi}/f_\varphi) \left(1 - \frac{p^2 \chi_0(t)^2}{4 {\tilde f}_\chi^2}\right),
\end{align}
where ${\tilde f}_\chi = \frac{f_\chi}{1 - q \kappa f_\chi/(p f_\varphi)}$. We see that the effect of the dark matter is always to {\em decrease} the magnitude of the gradient of the $\varphi$ potential, and this is always a subdominant effect when $\chi$ is near the bottom of its potential. In particular, the third condition of our fine-tuning argument, ``the part of the $\varphi$ potential that depends on the density of $\chi$ dominates the dynamics,'' never holds in this simple model.

We can go a step further and include a {\em third} term in the potential:
\begin{equation}
    \mathcal{L}_1 = \mathcal{L}_0 + \Lambda_3^4 \cos(r\,\chi/f_\chi + s\,\varphi/f_\varphi + \vartheta), \quad r, s \in \mathbb{Z},
\end{equation}
where $\vartheta$ is a constant phase offset, which can no longer be removed by a field redefinition. In this case, we see that
\begin{align}
    \frac{\partial V}{\partial {\tilde \varphi}} &= q \frac{\Lambda_2^4}{f_\varphi} \sin(q\, {\tilde \varphi}/f_\varphi) \langle \cos(p\,\chi/{\tilde f}_\chi)\rangle + s \frac{\Lambda_3^4}{f_\varphi} \sin(s \,{\tilde \varphi}/f_\varphi + \vartheta) \langle \cos(r\,\chi/{\hat  f}_\chi)\rangle \nonumber \\
    &\approx \frac{1}{f_\varphi} \left(U_0 - U_1 \frac{\chi_0(t)^2}{4 f_\chi^2}\right),
\end{align}
where now ${\hat f}_\chi = \frac{f_\chi}{1 - s \kappa f_\chi/(r f_\varphi)}$, and we have introduced the notation
\begin{align}
    U_0 &= q \Lambda_2^4 \sin(q \,{\tilde \varphi}/f_\varphi) + s \Lambda_3^4 \sin(s \,{\tilde \varphi}/f_\varphi + \vartheta), \nonumber \\
    U_1 &=  q p^2\frac{f_\chi^2}{{\tilde f}_\chi^2} \Lambda_2^4 \sin(q \,{\tilde \varphi}/f_\varphi)  + s r^2 \frac{f_\chi^2}{{\hat f}_\chi^2}\Lambda_3^4 \sin(s\, {\tilde \varphi}/f_\varphi + \vartheta).
\end{align}
From these expressions, we see that it is now {\em possible} that the $\chi$-dependent correction dominates over the $\chi$-independent potential, but this involves a fine-tuning. For example, if $\vartheta = \pi$ and $q \Lambda_2^4 \approx s \Lambda_3^4$, then $U_0$ approximately cancels. However, the additional parameters $p, r,$ and $\kappa$ allow for $U_1$ to not have a cancellation in this case.

We conclude that, although the Coleman-Weinberg argument of Sec.~\ref{subsec:naturalnessproblem} allowed for the possibility that a sufficiently light dark matter field $\chi$ could evade the one-loop fine-tuning problem, the most straightforward attempt to build a model evading this problem runs directly into a {\em tree-level} fine-tuning problem. This is one reason that we have focused, instead, on models where the one-loop Coleman-Weinberg potential is (mostly) canceled by summing over a monodromy tower.

Motivated by similar considerations of the Coleman-Weinberg effective potential,~\cite{DAmico:2016jbm} also considered a scenario in which both dark matter and dark energy arise from axion fields and interact with each other. They considered a case with {\em three} fields: a relatively heavy axion that serves as most of the dark matter, and two lighter and highly mixed axions, one of which behaves as dark energy and one of which behaves as a subdominant component of dark energy that only begins to oscillate at late times (see also~\cite{Amendola:2005ad,Hlozek:2014lca}). This evades the argument that we have given here, since the dominant dark matter in their scenario does not interact with the dark energy field. Their scenario is natural and leads to an interesting effective interaction between dark matter and dark energy. In this paper, we focus on the scenario with only one heavy species of dark matter.

\section{Dyon mass in the presence of massive fermions}\label{app:dyon_mass_formula}

In this section, we show that the dyon mass formula is unchanged in the heavy-fermion limit considered in this work. In the BPS limit, the dyon mass is determined solely by the electrostatic energy induced by the theta angle,
\begin{equation}
    m_D = \sqrt{m_{\rm M}^2+m_W^2\left(n-\frac{\theta}{2\pi}\right)^2}.
\end{equation}
However, the charge cloud around monopoles will be shielded by the wave function of massive fermions if there is matter content charged under the dark SU(2) group. Then we need to recompute the dyon mass in the presence of fermionic excitations, which are essentially dyonic bound states \cite{Hook:2024vhf}. To estimate the dyon mass term, we focus on the $J=0$ state of fermions, as higher angular momentum states should be treated as excited states of dyons. Then the 4-D problem reduces to a 2-D problem, and we can bosonize the 2-D fermions. The solitons of the bosonized theory map onto the original fermions, and after integrating out the photon the Lagrangian reduces to
\begin{equation}
    4\pi\mathcal{L}=\frac{1}{2}\sum_{i=1,2} (\partial \phi_i)^2+\left(\frac{\pi\, m_f}{2}\right)^2\sum_{i=1,2}\cos(\phi_i)-\frac{\alpha_D}{2\pi\, r^2}\left(\sum_{i=1,2}\frac{1}{2}\phi_i-\theta\right)^2,
\end{equation}
where $i=1,2$ represents different bosonic fields and we work with the equal mass case for the fermions.
Primes in this appendix denote derivatives with respect to $r$. The equation of motion is 
\begin{equation}
    \partial_r^2\phi_i=\left(\frac{\pi\,m_f}{2}\right)^2\sin(\phi_i)+\frac{\alpha_D}{4\pi\,r^2}(\sum_j\phi_j-2\theta).
    \end{equation}
The finite energy condition requires $\phi_i(\infty)=2\pi n_i$ and $\phi_1(0)+\phi_2(0)=2\theta$. The total charge of the system is 
\begin{equation}
    Q_{\rm EM} = \frac{n_1+n_2}{2}-\frac{\theta}{2\pi}.
\end{equation}
The Hamiltonian of the system is
\begin{equation}
    H=\frac{1}{4\pi}\int_0^\infty dr \left( \frac{1}{2}\sum_{i=1,2}\phi^{\prime 2}_i +\frac{1}{2}(\pi m_f)^2\sum_{i=1,2}\sin^2\frac{\phi_i}{2}+\frac{\alpha_D}{8\pi r^2}(\sum_{i=1,2} \phi_i-2\theta)^2\right).
\end{equation}
Now we will work with $\alpha_D\approx 0$ limit to analytically derive the approximate form of the energy. The static equation of motion becomes
\begin{equation}
    \phi_i''=\frac{\pi^2m_f^2}{4}\sin\phi_i.
\end{equation}
Multiplying both sides by $\phi_i^\prime$, we will find $(\phi_i')^2/2=-(\pi^2m_f^2\cos\phi_i) /4+C$. 
Note that the divergent Coulomb term should be regulated by the monopole profile if no light fermions are introduced. In the BPS limit, this corresponds to
\begin{equation}
    \delta E_{\rm EM}=\frac{1}{4\pi}\int_{1/(\alpha_Dm_{\rm M})}^\infty dr  \frac{\alpha_D}{8\pi r^2}\left(\sum_{i=1,2} \phi_i-2\theta\right)^2\approx m_{\rm M}\frac{\alpha_D^2}{2}\left(n-\frac{\theta}{2\pi}\right)^2.
\end{equation}
In the light fermion case ($m_f\ll \alpha_Dm_{\rm M}$), the fermions can screen the charge cloud and provide an even lower energy configuration. In such cases. the boundary condition changes to $\phi_i(\infty)=2\pi n_i$ and $\phi_i'(\infty)=0$. Later, we will obtain the dyon mass formula in the light fermion case based on this boundary condition, and explicitly see how it breaks down when it is preferred to excite the Coulomb energy itself instead of the solitonic fermion profile.
From the boundary condition $\phi_i(\infty)=2\pi n_i$ and $\phi_i'(\infty)=0$, we can determine $C=\pi^2m_f^2/4$ and 
\begin{equation}
    (\phi_i')^2=(\pi\, m_f)^2\sin^2\frac{\phi_i}{2}.
\end{equation}
Therefore, the mechanical part of the dyon mass is
\begin{equation}
    E_q=\frac{1}{4\pi}\int_0^\infty dr \sum_{i=1,2}\phi_i'^2=\frac{1}{4\pi} \sum_{i=1,2}\int_{\phi_i(0)}^{\phi_i(\infty)}d\phi_i|\phi_i'|.
\end{equation}
For solutions that do not wrap around, $\phi_i$ stays in a single branch where $\sin(\phi_i/2)\ge 0$, so $|\phi_i'|=\pi\,m_f\,\sin(\phi_i/2)$.
For simplicity, we assume $n_1=n_2=n$. Then the integral gives the compact result
\begin{equation}
    E_q = 2\,m_f\,\sin^2\left(\frac{\pi}{2}Q_{\rm EM}\right).
\end{equation}
The corrections from the electrostatic part can be estimated with a profile of $\phi_i(r)$ that is $\phi_i(r<1/m_f)\approx \theta$ and $\phi_i(r>1/m_f)\approx 2\pi n_i$.  We can calculate this contribution through
\begin{equation}
    E_q^{\rm EM} \approx \frac{1}{4\pi} \int_{1/m_f}^\infty dr \, \frac{\alpha_D}{8\pi r^2}( 2\pi (n_1+n_2)-2\theta)^2=\frac{\alpha
    _D}{2}m_f\,Q_{\rm EM}^2.
\end{equation}
This completes the analytical approximation of the dyon mass, which is $m_D=m_{\rm M}+E_q+E_q^{\rm EM}$, which gives
\begin{equation}\label{eq:dyon_mass_fermion}
m_D(\theta) \approx m_{\rm M} + 2m_f \sin^2\left[ \frac{\pi}{2} \left( n - \frac{\theta}{2\pi} \right) \right] + \frac{\alpha_D}{2} m_f \left( n - \frac{\theta}{2\pi} \right)^2.
\end{equation}
In the massless fermion limit, the dyon mass will reproduce the monopole mass without electric charge.
The above treatment is only a good approximation when the monopole radius $1/m_W$ is much smaller than the fermion clound $1/m_f$. More generally, the boundary condition used here is only valid when the excitation of the bosonized field is energetically favored, which gives the following condition
\begin{equation}
     2m_f \sin^2\left[ \frac{\pi}{2} \left( n - \frac{\theta}{2\pi} \right) \right] + \frac{\alpha_D}{2} m_f \left( n - \frac{\theta}{2\pi} \right)^2<  m_{\rm M} \frac{\alpha_D^2}{2} \left(n - \frac{\theta}{2\pi}\right)^2.
\end{equation}
Therefore, the effect of light fermions only starts to impact the dyon mass when $m_f\lesssim m_{\rm M} {\alpha_D^2}/{\pi^2}$. For the heavy fermion case, the original dyon mass formula without fermions will still apply, which is the relevant parameter space throughout this work.

\bibliographystyle{JHEP}
\bibliography{monopole_axion}

\end{document}